\documentclass{iopart}
\usepackage{psfrag,epsfig} 
\usepackage{iopams}

\renewcommand{\Re}{\mathrm{Re}\,}
\renewcommand{\Im}{\mathrm{Im}\,}
\newcommand{\rmu}{\mathrm{u}}

\newcommand{\eps}{\varepsilon}
\newcommand{\beps}{\boldsymbol\varepsilon}
\newcommand{\bbeps}{\boldsymbol{\bar\varepsilon}}

\newcommand{\ket}[1]{|#1\rangle}
\newcommand{\bra}[1]{\langle#1|}
\newcommand{\mv}[1]{\left\langle#1\right\rangle}
\newcommand{\ps}[2]{\bi{#1}\cdot\bi{#2}}
\newcommand{\cg}[2]{\bra{#1}#2\rangle} 
\newcommand{\eq}[1]{\begin{equation} #1 \end{equation}}
\newcommand{\eqlab}[2]{\begin{equation} #1
	\label{#2.eq}\end{equation}}
\newcommand{\refig}[1]{figure~\textup{\ref{#1.fig}}}
\newcommand{\refeq}[1]{\textup{(\ref{#1.eq})}}
\newcommand{\reftab}[1]{table~\textup{\ref{#1.tab}}}

\newcommand{\Je}{J_{\mathrm e}}
\newcommand{\me}{m_{\mathrm e}}
\newcommand{\Jg}{J}%_{\mathrm g}}
\newcommand{\mg}{m}%_{\mathrm g}}
\newcommand{\sfA}{\mathsf{A}}
\newcommand{\sfG}{\mathsf{G}}
\newcommand{\sfI}{\mathsf{I}}
\newcommand{\sfL}{\mathsf{L}}

\newcommand{\sfX}{\mathsf{X}}
\newcommand{\sfT}{\mathsf{T}}

\newcommand{\sfid}{\mathsf{1}}
\newcommand{\diff}{{\mathrm d}}

\newcommand{\sixj}[6]{\left\{\begin{array}{ccc}
		#1	& #2	& #3\\
		#4	& #5	& #6
		\end{array}\right\}}
\newcommand{\ninej}[9]{\left\{\begin{array}{ccc}
		#1	& #2	& #3\\
		#4	& #5	& #6\\
		#7	& #8	& #9
		\end{array}\right\}}

\newlength{\argwidth}
\newlength{\vertexheight}
\newcommand{\go}{\linethickness{0.085em}
   \begin{picture}(25,10)
        \put(0,3){\line(1,0){25}}
   \end{picture}}

\newcommand{\avg}{\linethickness{0.255em}
   \begin{picture}(25,6)
        \put(0,3){\line(1,0){25}}
   \end{picture}}

\newlength{\subwidth}

\newcommand{\connect}{%
	 \begin{picture}(0,20)(1,0)
	 \qbezier[30](0,7.5)(30,27.5)(60,7.5) 
         \end{picture}}

\newcommand{\self}{%
	 \begin{picture}(8,8)(0,0)
         \put(4,3){\circle*{8}}
         \end{picture}}

\newcommand{\lstep}{%
		\settowidth{\argwidth}{${\otimes}$}%
		\makebox[0.75\argwidth]{%
		\raisebox{\vertexheight}[2em][1.5em]{%
	 	\begin{picture}(0,25)(0,0)
			\put(0,0){\makebox(0,0){$\otimes$}}
			\put(0,25){\makebox(0,0){$\otimes$}}
			\qbezier[8](0,5)(0,12.5)(0,20)
		\end{picture}}}}

\newcommand{\rightarglstep}[2]{
	\raisebox{\vertexheight}[2em][1.5em]{%
        \begin{picture}(25,25)(0,0)
		\put(25,0){\vector(-1,0){8.5}}
		\put(16.5,25){\vector(1,0){8.5}}
		\put(12.5,0){\makebox(0,0){$\otimes$}}
		\put(12.5,25){\makebox(0,0){$\otimes$}}
		\qbezier[8](12.5,5)(12.5,12.5)(12.5,20)
		\put(25,25){\raisebox{-0.5ex}{$#1$}}
		\put(25,0){\raisebox{-0.5ex}{$#2$}}
   	\end{picture}}
	\settowidth{\argwidth}{$#2$}
	\hspace{\argwidth}
}
\newcommand{\leftarglstep}[2]{
	\settowidth{\argwidth}{$#1$}
	\hspace{\argwidth}
	\raisebox{\vertexheight}[2em][1.5em]{%
        \begin{picture}(25,25)(0,0)
        	\put(8.5,0){\vector(-1,0){8.5}}
        	\put(0,25){\vector(1,0){8.5}}
		\put(12.5,0){\makebox(0,0){$\otimes$}}
		\put(12.5,25){\makebox(0,0){$\otimes$}}
		\qbezier[8](12.5,5)(12.5,12.5)(12.5,20)
		\put(0,25){\hspace{-\argwidth}\raisebox{-0.5ex}{$#1$}}
		\settowidth{\argwidth}{$#2$}	
		\put(0,0){\hspace{-\argwidth}\raisebox{-0.5ex}{$#2$}}
   	\end{picture}}
}

\newcommand{\avgintens}{%
	\raisebox{\vertexheight}[2em][1.5em]{%
        \begin{picture}(25,25)(0,0)
		\linethickness{0.255em}
		\put(0,0){\line(1,0){25}}
		\put(0,25){\line(1,0){25}}
	\end{picture}}}

\newcommand{\leftargavgintens}[2]{%
	\settowidth{\argwidth}{$#2\,$}
	\hspace{\argwidth}
	\raisebox{\vertexheight}[2em][1.5em]{%
        \begin{picture}(25,25)(0,0)
		\put(8.5,0){\hspace{-\argwidth}\raisebox{-0.5ex}{$#2$}}
		\settowidth{\argwidth}{$#1\,$}
		\put(8.5,25){\hspace{-\argwidth}\raisebox{-0.5ex}{$#1$}}
		\linethickness{0.255em}
		\put(8.5,0){\line(1,0){25}}
		\put(8.5,25){\line(1,0){25}}
	\end{picture}}}

\newlength{\intensboxheight}
\newcommand{\intensbox}[1]{%
	\setlength{\intensboxheight}{1.3125em}
	\setlength{\fboxsep}{0.097em}
	\framebox{%
          \raisebox{0pt}[\intensboxheight][-\vertexheight]{$\ #1\ $}}}

\newcommand{\crossed}{%
	\raisebox{\vertexheight}[2em][1.5em]{%
        \begin{picture}(40,25)(0,0)
 	     	\qbezier[12](9,4)(20,12.5)(31,21)
      		\qbezier[12](9,21)(20,12.5)(31,4)
		\put(5,0){\makebox(0,0){$\otimes$}}
		\put(5,25){\makebox(0,0){$\otimes$}}
		\put(35,0){\makebox(0,0){$\otimes$}}
		\put(35,25){\makebox(0,0){$\otimes$}}
		\linethickness{0.255em}
		\put(8.5,0){\line(1,0){22.5}}
		\put(8.5,25){\line(1,0){22.5}}
    	\end{picture}}}

\newcommand{\argcrossed}[4]{
	\settowidth{\argwidth}{$#1$}
	\hspace{\argwidth}
	\raisebox{\vertexheight}[2em][1.5em]{%
        \begin{picture}(65,25)(0,0)
        	\put(8.5,0){\vector(-1,0){8.5}}
        	\put(0,25){\vector(1,0){8.5}}
		\put(55,0){\vector(-1,0){8.5}}
		\put(46.5,25){\vector(1,0){8.5}}
		\qbezier[12](16.5,4)(27,12.5)(38,21)
      		\qbezier[12](16.5,21)(27,12.5)(38,4)
		\put(12.5,0){\makebox(0,0){$\otimes$}}
		\put(12.5,25){\makebox(0,0){$\otimes$}}
		\put(42.5,0){\makebox(0,0){$\otimes$}}
		\put(42.5,25){\makebox(0,0){$\otimes$}}
		\linethickness{0.255em}
		\put(16,0){\line(1,0){22.5}}
		\put(16,25){\line(1,0){22.5}}
		\put(0,25){\hspace{-\argwidth}\raisebox{-0.5ex}{$#1$}}
		\put(55,25){\raisebox{-0.5ex}{$#2$}}
		\put(55,0){\raisebox{-0.5ex}{$#3$}}
		\settowidth{\argwidth}{$#4$}	
		\put(0,0){\hspace{-\argwidth}\raisebox{-0.5ex}{$#4$}}
   	\end{picture}}
	\settowidth{\argwidth}{$#3$}
	\hspace{\argwidth}
}

\newcommand{\triplecrossed}{%
	\raisebox{\vertexheight}[2em][1.5em]{%
          \begin{picture}(70,25)(0,0)
		\put(5,0){\makebox(0,0){$\otimes$}} 
		\put(5,25){\makebox(0,0){$\otimes$}} 
		\put(35,0){\makebox(0,0){$\otimes$}} 
		\put(35,25){\makebox(0,0){$\otimes$}} 
		\put(65,0){\makebox(0,0){$\otimes$}} 
		\put(65,25){\makebox(0,0){$\otimes$}} 
	        \qbezier[20](9,4)(35,12.5)(61,21)
		\qbezier[20](9,21)(35,12.5)(61,4)
		\qbezier[8](35,5)(35,12.5)(35,20)		
		\linethickness{0.255em}
		\put(8.5,0){\line(1,0){22.5}}
		\put(8.5,25){\line(1,0){22.5}}
		\put(38.5,0){\line(1,0){22.5}}
		\put(38.5,25){\line(1,0){22.5}}
		\end{picture}}}

\newcommand{\tdiagram}{%
	\raisebox{\vertexheight}[2.5em][1.5em]{%
         \begin{picture}(70,25)(0,0)
		\put(35,0){\makebox(0,0){$\otimes$}}
		\put(5,25){\makebox(0,0){$\otimes$}}
		\put(35,25){\makebox(0,0){$\otimes$}}
		\put(65,25){\makebox(0,0){$\otimes$}}
		\qbezier[8](35,5)(35,12.5)(35,20)
		\qbezier[30](9,29.5)(34.5,49.5)(60,29.5)
		\linethickness{0.255em}
		\put(8.5,25){\line(1,0){22.5}}
		\put(38.5,25){\line(1,0){22.5}}	
		\end{picture}}}

\newcommand{\tdiagramcc}{%
	\raisebox{\vertexheight}[2em][2em]{%
         \begin{picture}(70,25)(0,0)
		\put(35,25){\makebox(0,0){$\otimes$}}
		\put(5,0){\makebox(0,0){$\otimes$}}
		\put(35,0){\makebox(0,0){$\otimes$}}
		\put(65,0){\makebox(0,0){$\otimes$}}
		\qbezier[8](35,5)(35,12.5)(35,20)
		\qbezier[30](9,-4.5)(34.5,-24.5)(60,-4.5)
		\linethickness{0.255em}
		\put(8.5,0){\line(1,0){22.5}}
		\put(38.5,0){\line(1,0){22.5}}	
		\end{picture}}}

\newcommand{\vertex}[4]{
	\settowidth{\argwidth}{$#1$}
	\hspace{\argwidth}
	\raisebox{\vertexheight}[2em][1.5em]{%
        \begin{picture}(25,25)(0,0)
        	\put(0,0){\line(1,0){25}}
        	\put(0,25){\line(1,0){25}}
        	\put(8.5,0){\line(0,1){25}}
        	\put(16.5,0){\line(0,1){25}}
		\put(0,25){\hspace{-\argwidth}\raisebox{-0.5ex}{$#1$}}
		\put(25,25){\raisebox{-0.5ex}{$#2$}}
		\put(25,0){\raisebox{-0.5ex}{$#3$}}
		\settowidth{\argwidth}{$#4$}	
		\put(0,0){\hspace{-\argwidth}\raisebox{-0.5ex}{$#4$}}
   	\end{picture}}
	\settowidth{\argwidth}{$#3$}
	\hspace{\argwidth}
}

\newcommand{\horizontal}[4]{
	\settowidth{\argwidth}{$#1$}
	\hspace{\argwidth}
	\raisebox{\vertexheight}[2em][1.5em]{%
        \begin{picture}(15,25)(0,0)
        	\put(0,0){\line(1,0){15}}
        	\put(0,25){\line(1,0){15}}
		\put(0,25){\hspace{-\argwidth}\raisebox{-0.5ex}{$#1$}}
		\put(15,25){\raisebox{-0.5ex}{$#2$}}
		\put(15,0){\raisebox{-0.5ex}{$#3$}}
		\settowidth{\argwidth}{$#4$}	
		\put(0,0){\hspace{-\argwidth}\raisebox{-0.5ex}{$#4$}}
   	\end{picture}}
	\settowidth{\argwidth}{$#3$}
	\hspace{\argwidth}
}
\newcommand{\diagonal}[4]{
	\settowidth{\argwidth}{$#1$}
	\hspace{\argwidth}
	\raisebox{\vertexheight}[2em][1.5em]{%
        \begin{picture}(25,25)(0,0)
		\qbezier(0,2.5)(5.5,7)(11,11.5)
		\qbezier(25,22.5)(19.5,18)(14,13.5)
		\qbezier(0,22.5)(12.5,12.5)(25,2.5)
		\put(0,25){\hspace{-\argwidth}\raisebox{-0.5ex}{$#1$}}
		\put(25,25){\raisebox{-0.5ex}{$#2$}}
		\put(25,0){\raisebox{-0.5ex}{$#3$}}
		\settowidth{\argwidth}{$#4$}	
		\put(0,0){\hspace{-\argwidth}\raisebox{-0.5ex}{$#4$}}
   	\end{picture}}
	\settowidth{\argwidth}{$#3$}
	\hspace{\argwidth}
}
\newcommand{\vertical}[4]{
	\settowidth{\argwidth}{$#1$}
	\hspace{\argwidth}
	\raisebox{\vertexheight}[2em][1.5em]{%
        \begin{picture}(20,25)(0,0)
        	\put(0,7.5){\line(0,1){12.5}}
        	\put(20,7.5){\line(0,1){12.5}}
		\put(0,25){\hspace{-0.5\argwidth}\raisebox{-0.5ex}{$#1$}}
		\put(20,25){\hspace{-0.5\argwidth}\raisebox{-0.5ex}{$#2$}}
		\put(20,0){\hspace{-0.5\argwidth}\raisebox{-0.5ex}{$#3$}}
		\settowidth{\argwidth}{$#4$}	
		\put(0,0){\hspace{-0.5\argwidth}\raisebox{-0.5ex}{$#4$}}
   	\end{picture}}
	\settowidth{\argwidth}{$#3$}
	\hspace{\argwidth}
}

\newcommand{\twistedvertex}[4]{
	\settowidth{\argwidth}{$#1$}
	\hspace{\argwidth}
	\raisebox{\vertexheight}[2em][1.5em]{%
        \begin{picture}(25,25)(0,0)
        	\put(0,0){\line(1,0){25}}
        	\put(0,25){\line(1,0){25}}
	        \qbezier(8.5,0)(8.5,6.25)(12.5,12.5)
		\qbezier(12.5,12.5)(16.5,18.75)(16.5,25)
        	\qbezier(8.5,25)(8.5,18.75)(11.86,13.5)
		\qbezier(13.14,11.5)(16.5,6.25)(16.5,0)
 		\put(0,25){\hspace{-\argwidth}\raisebox{-0.5ex}{$#1$}}
		\put(25,25){\raisebox{-0.5ex}{$#2$}}
		\put(25,0){\raisebox{-0.5ex}{$#3$}}
		\settowidth{\argwidth}{$#4$}	
		\put(0,0){\hspace{-\argwidth}\raisebox{-0.5ex}{$#4$}}
   	\end{picture}}
	\settowidth{\argwidth}{$#3$}
	\hspace{\argwidth}
}

\begin{document}

\title{Multiple scattering of light by atoms with
internal degeneracy}
\author{Cord Axel M\"{u}ller$^{1,2}$ and 
Christian Miniatura$^1$}
\address{1  Laboratoire Ondes et D\'esordre, FRE 2302 du CNRS,
  1361 route des Lucioles,\\
  F-06560\ Valbonne, France}
\address{2 Max-Planck-Institut f\"ur Physik komplexer Systeme, 
  N\"othnitzer Str. 38, \\
  D-01187  Dresden, Germany}

\ead{\mailto{cord@mpipks-dresden.mpg.de}, \mailto{miniat@inln.cnrs.fr}}

\date{\today}

\begin{abstract}
An analytical microscopic theory for the resonant multiple scattering
of light by cold atoms with arbitrary internal degeneracy is
presented. It permits to calculate the average amplitude and the average
intensity for one-photon states of the full transverse electromagnetic
field in a dilute medium of unpolarized atoms. 
Special emphasis is laid upon an 
analysis in terms of irreducible representations
of the rotation group. 
It allows to sum explicitly the
ladder and maximally crossed diagrams, giving the average intensity in
the Boltzmann 
approximation and the interference corrections responsible for weak
localization and coherent backscattering. 
The exact decomposition into field modes shows that the
atomic internal degeneracy contributes to the depolarization of the average
intensity and suppresses the interference 
corrections. 
Static as well as dynamic quantities like the transport velocity,
diffusion constants  
and relaxation times for all field modes and all atomic transitions
are derived.      

\end{abstract}

%\pacs{32.80.-t,05.60.Gg,72.15.Rn}
%Quantum transport
%Photon interactions with atoms
%Localization effects (Anderson or weak localization)
%\submitto{\JPA}

\section{Introduction}

Wave propagation in disordered media has been an active field of
research for more than a century. The first descriptions of electronic
conduction in weakly disordered metals by Drude \cite{Drude00} and of light
propagation in interstellar clouds by Schuster \cite{Schuster05} were
based on Boltzmann-type transport equations. Despite their 
successful predictions 
(\emph{e.g.} Ohm's law), 
these theories overlooked the subtle role of 
interference. 
P. Anderson first showed that sufficiently strong disorder can induce 
a metal-insulator transition for quantum-mechanical
wave functions on a random lattice, 
a phenomenon baptized
Anderson localization \cite{Anderson58}. 
Furthermore, it was realized that interference influences 
transport even far from the localization transition 
and that also other waves are concerned.  
 For example, the constructive interference of
counter-propagating light amplitudes in the backward
direction, as first pointed out by Watson \cite{Watson69}, survives 
an average over disorder, giving
rise to the coherent backscattering (CBS) peak
\cite{Tsang84,Akkermans88,vanderMark88}. In the past 20 years, the
field of mesoscopic physics has developed, with  
beautiful experimental and theoretical results pertaining to the weak
localization regime  
\cite{Bergmann84,Lee85,Houches94}.

Atoms, as natural realizations of highly resonant 
identical point 
scatterers, have been proposed as an ideal sample for multiple light
scattering and even localization 
\cite{Sornette88,Nieuwenhuizen94}.  
However, as first discovered in experiments on CBS by rubidium atoms 
\cite{Labeyrie99,Labeyrie00} and recently 
confirmed on strontium atoms \cite{Bidel}, 
the quantum internal structure of atomic scatterers has a strong
impact on light transport properties. The Zeeman 
degeneracy of the probed dipole transition 
reduces the   
interference contrast of coherent backscattering \cite{Jonckheere00}
dramatically. The single and double
scattering contribution to the CBS peak for atoms with arbitrary
internal degeneracy have been calculated in \cite{Mueller01}. 

In the present contribution, we develop an analytical theory of
multiple resonant  
one-photon elastic scattering in an infinite medium consisting of a 
dilute gas of cold atoms at fixed 
classical positions, but with
arbitrary internal degeneracy.   
Our results are a first step towards a microscopic 
generalization of the existing
theories for the scattering of transverse vector waves by Rayleigh
point scatterers \cite{Ozrin92a,Amic97} on the one hand and of scalar waves by
anisotropic scatterers \cite{Amic96,Ozrin92b} on the other.  
The paper is organized as follows. In section \ref{Hamiltonian}, 
we define the Hamiltonian for the coupled system light and atoms. 
In section \ref{amplitude.sec}, we calculate the average propagator of
a one-photon Fock state inside the scattering medium with the aid of the Dyson
equation and the photon self-energy. 
In section \ref{intensity.sec}, we define the average intensity
for the full vector field in terms of photo-detection probability. 
 The corresponding Bethe-Salpeter equation 
is then solved by evaluating the so-called 
irreducible vertex in the Boltzmann approximation. 
Technically, this is
achieved by a systematic analysis in terms of irreducible tensor
operators. 
In section \ref{sum.sec}, 
we sum the so-called ladder series in the static limit 
by applying the same method of irreducible tensors to 
the intensity propagator of transverse field modes.  
The interference correction described by 
the maximally crossed diagrams is then easily obtained by applying  
substitution rules. In 
section \ref{dyn.sec}, we generalize our results to the dynamic case and 
derive analytical expressions for transport velocities, diffusion
constants, extinction lengths and relaxation times for the 
different field modes and all atomic transitions. We conclude by possible 
extensions of the work.

\section{Description of the system}
\label{Hamiltonian}

\subsection{Hamiltonian}

In order to analyze the role of the atomic 
quantum internal structure, we must treat the atomic 
internal degrees of freedom 
quantum mechanically. For reasons of symmetry, 
the light field is treated quantum mechanically as well. 
Using natural units such that $\hbar=c=1$, the global system 
``matter + light'' is then described by the Hamiltonian $H=H_0 + V$ where 
\eqlab{
H_0 = H_{\mathrm{em}} + H_{\mathrm{at}} = \sum_{{\bf k}, \beps \perp \bi k} \omega_{\bi k} \: 
	a_{\bi k\beps}^\dagger a_{\bi k\beps} + 
\sum_{\alpha=1}^{N} H_{\mathrm{at}}(\alpha).
}{H_0}
$H_{\mathrm{em}}$ is the Hamiltonian of the 
free electromagnetic field.  
The annihilation and creation operators for the field modes
(wave vector $\bi k$, transverse polarization $\beps$, free dispersion
relation $\omega_{\bi k} = k$)  
obey the usual bosonic commutation relation 
$[ a_{{\bi k}\beps},a_{{\bi k'}\beps'}^\dagger] =
\delta_{\beps,\beps'} \delta_{\bi k,\bi k'}$. 
For brevity, a one-photon Fock state will be noted 
$\ket{\bi k\beps}\equiv a_{\bi k\beps}^\dagger\ket{0}$ where 
the transversality $\beps\cdot\bi k=0$ is understood.
$H_{\mathrm{at}}$ is the sum of the individual internal atomic Hamiltonians 
\eq{
 H_{\mathrm{at}}(\alpha) = \omega_0 
\sum_{\me=-\Je}^{\Je}\ket{\Je\me(\alpha)}\bra{\Je\me(\alpha)}. 
}
Each atom $\alpha=1,\dots,N$ is described as a \emph{degenerate} closed 
two-level system with an energy separation 
$\omega_0$ between the ground-state with total angular momentum 
$\Jg$ and the excited-state 
with total angular momentum $\Je$ (\refig{transition}). These levels are  
resonantly coupled by the dipole interaction 
\eq{
V = \sum_{\alpha=1}^{N} V_\alpha = 
- \sum_{\alpha=1}^{N}\bi D_\alpha\cdot\bi E(\bi r_\alpha). 
}
The individual atomic dipole operator 
$\bi D_\alpha$ acts on the Hilbert space  
$\mathcal{H}_{\mathrm{g}}\oplus \mathcal{H}_{\mathrm{e}}$ of the 
internal atomic states of atom $\alpha$. 
Dropping subscript $\alpha$, we define the reduced dipole operator 
$\bi d=\bi D/d$ 
where $d=\bra{\Je}|\bi D|\ket{\Jg}/\sqrt{2\Je+1}$. The 
matrix elements of its spherical components 
$d_q\equiv \bi e_q\cdot\bi d $, $q=-1,0,+1$ are the Clebsch-Gordan
coefficients   
 $\bra{\Je\me}d_q\ket{\Jg\mg} = \cg{\Je\me}{\Jg1\mg q}$ 
according to the Wigner-Eckhart theorem \cite{Edmonds60}.  
In the dipole interaction, the electric field operator
\eqlab{
\bi E(\bi r) = \rmi \sum_{\bi k,\beps \perp \bi k}  \mathcal{E}_\omega 
   \left(\beps_{\bi k} a_{\bi k\beps} \rme^{\rmi\ps{k}{r}}  
 - \bbeps_{\bi k} a_{\bi k\beps}^\dagger
 \rme^{-\rmi\ps{k}{r}}\right)
}{E}
is evaluated at the center of mass $\bi r_\alpha$ of each scattering
atom. 
The field oscillator strength $\mathcal{E}_\omega =
(\omega/2\epsilon_0L^3)^{1/2}$ is given in terms of a quantization
volume $L^3$ that will disappear from physically relevant
expressions in the limit of an infinite medium by virtue of the rule    
$L^{-3}\sum_{\bi k} (\dots) \mapsto (2\pi)^{-3} \int \diff^3 k  (\dots)$. 

\begin{figure}
\begin{center}
\includegraphics{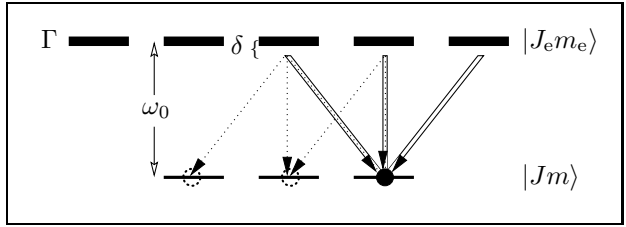}
\caption{Resonant degenerate dipole transition, here for $J=1$,
$\Je=2$. The atomic resonance frequency is noted $\omega_0$, the detuning of
the light probe $\delta=\omega-\omega_0$, and the natural width of the
excited atomic state $\Gamma$. The internal magnetic quantum numbers
$m$ are coupled to the polarization of scattered photons (full and
dotted lines).}
\label{transition.fig}
\end{center}
\end{figure}

\subsection{Length scales hierarchy}

The detailed description of wave propagation in a large sample 
of scatterers at quenched random positions is such a formidable task 
that one rather resorts to a statistical description, 
catching the generic features through configuration 
averaged quantities  
(average amplitude, average intensity, \emph{etc})
\cite{Lagendijk96,Frisch,vanRossum99}.  
In the following, we will consider the limit of a  
disordered \emph{infinite} medium where 
resonant point scatterers are distributed with constant spatial density $n$. 
The absence of boundaries greatly simplifies the theory since  
translational symmetry is restored on average. 
The importance of interference depends essentially 
on the hierarchy 
between the physically relevant length scales in the system, here the
wave length $\lambda$ and the mean inter-particle distance $n^{-1/3}$.
We will concentrate on the \emph{low density regime} 
where $\lambda \ll n^{-1/3}$. 
This inequality, alternatively rewritten as $n\lambda^3 \ll 1$, 
justifies a semi-classical 
description of propagation along rays between consecutive scatterers
that define scattering paths and their associated amplitudes.  
It also implies that scattering 
paths involving different scatterers are \emph{uncorrelated}: since
their phase difference  
greatly exceeds $2\pi$, the associated interference term can be
expected to average to zero. 
Also, recurrent scattering sequences 
(visiting a given scatterer more than once) can be neglected:  
this is the \emph{independent scattering approximation} (ISA)
\cite{Lagendijk96}. 
In this dilute regime, the elastic mean free path (mean distance 
traveled between two successive scattering events) is given by 
$\ell = 1/n\sigma$ where $\sigma$ is the total cross-section for
elastic scattering. 
Since we treat resonant point scatterers where $\sigma\sim\lambda^2$, 
the low density condition implies $\lambda \ll \ell$: 
wave scattering is described in the far field. On the same ground, the
equivalent relation $\sqrt{\sigma}\ll \ell$ shows that recurrent scattering is
indeed negligible. And finally, the low density condition implies $k\ell \gg 1$
where $k=2\pi/\lambda$, which defines the \emph{weak localization
regime}.

\section{Average light propagation}
\label{amplitude.sec}

The simplest quantity of interest for wave propagation 
in random media is the average amplitude, so we first compute the
configuration-averaged propagator $\mv{G(\omega)}$.

\subsection{The one-atom scattering operator}

In the weak localization regime, a scattered wave has reached its 
asymptotic limit when undergoing a subsequent collision.
 Hence a central 
quantity in the theory is the one-atom transition operator $T_\alpha
(z)$, given by the Born
series $T_\alpha(z)=V_\alpha+V_\alpha G_0(z)V_\alpha +\dots$  in powers of 
the interaction $V_\alpha$ and the
free resolvent operator $G_0(z)=(z-H_0)^{-1}$ \cite{Goldberger67}.
The scattering amplitude associated 
to the elastic scattering process 
$\ket{i}=\ket{\bi
k\beps ; \Jg\mg} \to \ket{f} = \ket{\bi k'\beps' ; \Jg\mg'}$ is 
proportional to 
\eqlab{
    \bra{f} T_\alpha(\omega+\rmi0) \ket{i}= 
   \bra{\Jg\mg'}\bbeps' \cdot \bi t(\omega)\cdot \beps \ket{\Jg\mg} \,
	\rme^{\rmi(\bi k-\bi k')\cdot \bi r_\alpha}
}{amplidet} 
where the notation 
$\omega+\rmi 0$ indicates that retarded
propagators are used in the Born series and that the matrix 
elements are taken on-shell 
($\omega = \omega_{\mathrm{\bi k}} = \omega_{\mathrm{\bi k'}}$). 
Note that the incident and scattered wave vector 
appear only in the exponential on the right hand side: in 
the dipole approximation, the atom indeed behaves as a point
scatterer, and the external and internal degrees of freedom are
factorized. 
The frequency dependence of the internal 
transition operator $\bi t(\omega)$ is 
\eq{
	t(\omega)=
	\frac{3}{2\pi\rho_0(\omega)}\frac{\Gamma/2}{\delta+\rmi\Gamma/2}
}
where $\Gamma= d^2\omega_0^3/3\pi\epsilon_0$ is the natural width of
the excited atomic state, $\delta= \omega -\omega_0$ is the detuning
of the probe light from the atomic resonance and
$\rho_0(\omega)=L^3\omega^2/2\pi^2$ is the free photon spectral
density.  
This form of a 
resonant scalar t-matrix is well known 
 in the context of classical point scatterers \cite{deVries98}.  
But as the atomic scatterer has an internal structure, particular
attention must be paid to the tensor character of the scattering
operator. For a given atomic transition
$\mg \mapsto \mg'$, the Cartesian elements of the t-matrix are 
\eqlab{
t_{ij}(\mg,\mg',\omega) = t(\omega) \bra{\Jg\mg'}d_i d_j\ket{\Jg\mg}. 
}{tmatrix} 
This $3\times 3$ t-matrix can be decomposed into its irreducible
components with respect to rotations, its scalar part (or trace), its
antisymmetric part and symmetric traceless part: 
\eqlab{
 t_{ij}  =  \underbrace{\frac{1}{3}\delta_{ij}t_{kk}}
		_{\displaystyle t_{ij}^{(0)}}  
+ \underbrace{\frac{1}{2}(t_{ij}-t_{ji})}
	_{\displaystyle t_{ij}^{(1)}}  
+ \underbrace{\frac{1}{2}(t_{ij}+t_{ji}) - \frac{1}{3}\delta_{ij}
t_{kk}}_{\displaystyle t_{ij}^{(2)}}
}{tdecomposition}
In the case of a point dipole  ($\Jg=0,\Je=1$), 
only the scalar part is non-zero, $t_{ij} =t (\omega) \delta_{ij}$. 
In the more general situation of arbitrary degeneracy, 
we  have to determine the influence of precisely  
the non-scalar parts on the average light evolution.  

\subsection{Averaging procedure}

In the statistical description of light evolution inside a 
disordered scattering
medium, configuration averaged quantities $\mv{\cdot}\equiv
\Tr[(\cdot)\rho_{\mathrm{at}}]$  
are obtained by tracing 
over the matter degrees of freedom.
The external degrees of freedom are the uncorrelated classical positions 
$\{\bi r_\alpha\}$, and the average is performed 
by spatial integration. The internal average is a trace over the 
internal density matrix. Since scattering theory relates asymptotically free 
states, this density matrix only describes the 
statistical properties of the ground state. We further assume the 
total density matrix $\rho_{\mathrm{int}} = \bigotimes_\alpha \rho_J(\alpha)$ 
to be a direct product of one-atom density 
matrices, each proportional to the ground state unit matrix 
(complete statistical mixture, no internal correlations):
\eqlab{
\rho_J(\alpha)=\frac{1}{2\Jg+1}\sum_{\mg} 
\ket{\Jg\mg(\alpha)}\bra{\Jg\mg(\alpha)}. 
}{rho0}

The free evolution of the photon state is then determined by the
resolvent operator  
\eqlab{
g_0(z) = \mv{G_0(z)} = (z-H_{\mathrm{em}})^{-1}  
}{g0}   
where the average of the free resolvent operator
$G_0(z)=(z-H_0)^{-1}$ (which is of course independent of the disorder) 
simply projects onto the atomic ground state chosen to have zero energy. 
The full average evolution in the presence of the scatterers is
described by the average resolvent operator $\mv{G(z)}$ which can be
expanded in the Born series
\eqlab{
\mv{G(z)} = g_0(z)+g_0(z)\mv{V} g_0(z) + g_0(z)\mv{V G_0(z) V} g_0(z) +
\dots
}{avG}
Summing all repeated interactions with the same scatterer, one obtains
a series in terms of the one-atom scattering operators $T_\alpha(z) =
V_\alpha+ V_\alpha G_0(z) V_\alpha +\dots$, 
\eqlab{\fl
\mv{G(z)} = g_0(z)+\sum_\alpha g_0(z) \mv{T_\alpha(z)} g_0(z) +
\sum_{\alpha\ne \beta}g_0(z) \mv{T_\alpha(z)} g_0(z) \mv{T_\beta(z)} g_0(z) +
\dots
}{avGdeT}
The average one-atom transition operator $\mv{T_\alpha(z)}$ does not depend 
on its index
$\alpha$ since all atoms are equally distributed, 
so that this series can be represented diagrammatically as 
\eqlab{
\avg= \go + N  \;\go {\otimes} \go 
+ N(N-1) \;\go {\otimes} \go {\otimes} \go 
+\dots}{avgdet}
The right hand side features the contributions of 
free evolution, single scattering,
double scattering, and so forth. 
Starting from the following term of triple scattering (not shown), this average
contains second- and higher-order moments of one-atom scattering
operators, arising from recurrent scattering by the
same scatterer.  
As usual in diagrammatic expansions \cite{Frisch}, one introduces an operator
representing the sum of irreducible contributions, the self-energy. 

\begin{table}
\caption{Definition of symbols in Feynman diagrams.}
\renewcommand{\arraystretch}{1.2}
\begin{indented}
\item[]\begin{tabular}{@{}clll}
\br
Symbol 	& Definition	& Name				&Equation	\\
\mr
$\go$	& $g_0(z)$ 	&free photon propagator		&\refeq{g0} 	\\
$\avg$	& $\mv{G(z)}$	&average photon propagator	&\refeq{avG} 	\\ 
$\otimes$&$\mv{T_\alpha(z)}$ 
			&average one-atom transition operator 
			&\refeq{avGdeT} \\
$\self$	& $\Sigma(z)$	&self-energy			&\refeq{dyson} 	\\
\dotted &		&link between identical scatterers &		\\
$\lstep$ &$\mv{\overline{T}_\alpha(z_1)T_\alpha(z_2)}$ 
			&average single-scattering vertex
			&\refeq{UdeTdaggerT} \\
$\vertex{i}{j}{k}{l}$ &$\sfI_{il;jk}$	
			&atomic polarization vertex
			&\refeq{boxdef} 	\\
$\twistedvertex{i}{j}{k}{l}$	&$\sfX_{il;jk}$	
			&crossed atomic polarization vertex
			&\refeq{vertextordu} 	\\
\br
\end{tabular}
\renewcommand{\arraystretch}{1.0}
\end{indented}
\end{table}		

\subsection{Self-energy and Dyson equation}

The average one-photon Green function $\mv{G(z)}$ 
satisfies the Dyson equation
\eqlab{
  \mv{G(z)} \equiv g_0(z) + g_0(z) \Sigma(z) \mv{G(z)}   
}{dyson}
that reads in diagrammatic form 
\eqlab{ 
\avg \equiv \go + \go \self \avg.
}{dysondiag}
By iteration, one recognizes a  
geometrical series which formally sums up to   
\eqlab{ 
\mv{G(z)} = [g_0(z)^{-1}-\Sigma(z)]^{-1}.
}{mvGdeSigma}
The Dyson equation actually defines the self-energy 
$\Sigma(z)$ whose exact calculation remains impossible in most cases. 
Nevertheless, introducing the self-energy 
has two major advantages: First, any approximate expression for 
the self-energy will yield an approximate, but \emph{non-perturbative} 
result for the average propagator 
\refeq{mvGdeSigma}. 
Second, 
the perturbative expansion of the self-energy in a power series 
and its truncation are controlled by the small parameter $n\lambda^3$. 
The self-energy contains precisely all 
irreducible diagrams, \emph{i.e.} those that cannot be separated into independent
diagrams by cutting a single line,   
\eqlab{
\Sigma (z) \equiv \self = N \otimes
+ N(N-1)\; {\otimes} {\connect} \go {\otimes} \go {\otimes}
+\dots 
}{selfenergie}
Here, the dotted line identifies the same scattering operator
appearing twice. 
Substituting this series into the Dyson equation 
\refeq{dysondiag} indeed reproduces
the average propagator \refeq{avgdet}  
in the thermodynamic limit $N,L\to \infty$ at constant density $n=N/L^3$.  

For a dilute medium $n\lambda^3\ll 1$, the independent scattering 
approximation (ISA) amounts to the first order truncation  
	$\Sigma (z)\approx N \otimes  =  N \mv{T_\alpha(z)}$.
With \refeq{amplidet}, the one-photon matrix elements of the
self-energy are  
\eqlab{ 
	\bra{\bi k'\beps'} \Sigma(\omega) \ket{\bi k\beps} = N
	\mv{\bbeps'\cdot \bi t(\omega) \cdot \beps}_{\mathrm{int}}  
   	\langle \rme^{\rmi(\bi k-\bi k')\cdot \bi r}\rangle_{\mathrm{ext}} . 
}{elmatSigma}
The external average gives 
$\langle \rme^{\rmi(\bi k-\bi k')\cdot \bi r}\rangle_{\mathrm{ext}} =
	\delta_{\bi k,\bi k'}$, 
indicating that
the self-energy is diagonal in momentum space 
as required by the statistical invariance under translations. 
The internal average of the
scattering operator  \refeq{tmatrix} with the scalar density matrix
\refeq{rho0} is elementary using the closure relation of
Clebsch-Gordan coefficients \cite{Edmonds60}, 
$\bra{\bi k'\beps'} \Sigma(\omega) \ket{\bi k\beps} = 
N M_J \, t(\omega) \, (\bbeps'\cdot\beps) \, \delta_{\bi k,\bi k'}$. 
The scalar product of polarization vectors, 
$\bbeps'\cdot\beps = \delta_{\beps,\beps'}$ shows that the self-energy,
as a second-rank tensor, is proportional to the unit matrix, 
$\boldsymbol\Sigma(\omega)=\Sigma(\omega) \boldsymbol 1$, 
as required by statistical invariance
under rotations. 
Note that   
\eqlab {\Sigma(\omega) = 
	n\,M_J\,\frac{3\pi}{\omega^2}\frac{\Gamma/2}{\delta+\rmi\Gamma/2}
}
{sigmadiag}
depends directly on the number density $n=N/L^3$ so that the
thermodynamic limit is trivial. For convenience, we define the 
multiplicity ratio $M_J = (2\Je+1)/3(2J+1)$ with the non-degenerate 
limit $M_0=1$. 

In optics and atomic physics, 
the response of an atom to an external electric
field is often expressed in terms of the atomic polarizability 
$\boldsymbol \alpha(\omega) = - 
\frac{2L^3}{\omega}\mv{\bi t(\omega)}$ \cite{Berestetski89}. 
In the ISA, the polarizability is  
simply proportional to the self-energy, 
$\Sigma(\omega) = -\frac{\omega}{2} n \alpha(\omega)$, and thus a
scalar. The dilute regime is alternatively characterized by $n |\alpha| \ll 1$.

\subsection{Effective medium}

Just like the free propagator and the self-energy, the average 
retarded photon propagator %\refeq{mvGdeSigma} 
is diagonal in momentum and
polarization: $\bra{\bi k'\beps'} \mv{G(\omega)} 
\ket{\bi k\beps} = \mv{G(k,\omega)} \, \delta_{\bi k,\bi k'} \, 
(\bbeps'\cdot\beps)$ with 
\eqlab{ 
	\mv{G(k,\omega)} = \frac{1}{\omega-k-\Sigma(\omega)}. 
}{Gdiag}
Its singularities are the solutions of the complex dispersion relation
$\omega-k-\Sigma(\omega)=0$. 
The average propagation can be seen to proceed through an effective 
medium with the \emph{complex-valued} frequency-dependent refractive index 
$n_r=k/\omega = 1-\Sigma(\omega)/\omega$. 
The imaginary part of the self-energy defines the elastic mean free
time 
\eqlab{
	\frac{1}{\tau} = -2 \Im\Sigma(\omega)  
}{mft}
of a pure one-photon state $\ket{\bi k\beps}$ inside the medium. 
Our perturbative approach proves to be consistent: the 
correction $1/\tau$ to the free frequency $\omega$
is small by definition in the weak 
scattering regime $1/\omega\tau\ll 1$.  
A propagating wave-packet is therefore exponentially
damped on average with a scattering mean free
path $\ell = \tau$ (remember $c=1$). 
This depletion is not caused by absorption, but
by elastic collisions into initially empty field states. 
At low densities, and by virtue of the 
optical theorem, the mean free path is simply 
$\ell = 1/n\sigma$ in terms of the total elastic scattering cross
section 
\eqlab{\sigma 
       = - 2 L^3 \,\Im\bra{\bi
	k\beps}\mv{T(\omega)}\ket{\bi k\beps} = 
       M_J\,\frac{6\pi}{k^2}\frac{1}{1+4\delta^2/\Gamma^2}.
}{sectionefficace}
In all of the above expressions, only a modest dependence on $\Jg$
and $\Je$ arises through the multiplicity ratio  $M_J$.   
Obviously, the quantum
internal structure has almost no impact on the average amplitude:
under an average over the 
scalar density matrix \refeq{rho0}, only the scalar part or trace of the
t-matrix \refeq{tdecomposition} can survive. 
We are left with a scalar theory for the average
amplitude, describable in terms of the polarizability alone. 
But one should not conclude prematurely that the internal
structure has no impact on the average intensity which, of course,  
must be carefully distinguished from the square of the average
amplitude.   

%%%%%%%%%%%%%%%%%%%%%%%%%%%%%%%%%%%%%%%%%%%%%%%%%%%%
\section{Average light intensity}
\label{intensity.sec}

In order to determine 
the average population of initially empty field modes, 
we now turn to the average intensity, defined in terms of 
photo-detection probability. 

\subsection{Intensity propagation kernel}

Since the total Hamiltonian $H=H_0+V$ is time-independent, the 
density matrix $\rho$ of the coupled system ``atoms +
field'' evolves according to $\rho(t) = U(t) \rho\, U^\dagger(t)$
where the forward time evolution operator $U(t)$ is the Fourier 
transform of the retarded propagator $G(\omega)=(\omega-H+\rmi 0)^{-1}$,     
\eqlab{
U(t)=-\frac{1}{2\pi\rmi}\int_{-\infty}^\infty \rmd\omega\, G(\omega) 
\rme^{-\rmi \omega t}, \quad t>0.  
}{U}
The measurable average light intensity at position $\bi r$ and time $t$ is
proportional to the average photo-detection probability \cite{CCTrouge},
\eqlab{
\mv{I(\bi r,t)}= \mathcal{N} \, \Tr[\rho(t) \bi E^{(-)}(\bi r) \cdot 
\bi E^{(+)}(\bi r)].  
}{I}
Here, the factor $\mathcal{N}$ contains the detection efficiency, and 
$\bi E^{(\pm)}(\bi r)$ are the annihilation and creation 
components, respectively, of the electric field operator
\refeq{E}. Their normal ordering assures that the photon vacuum state 
yields zero
intensity. 
Using \refeq{U}, we define the Fourier
transform of the intensity 
%\eq{
$	\mv{I(\bi r,t)} = L^{-3} \sum_{\bi q} (2\pi)^{-2} \int 
              \rmd\omega  \rmd\Omega 
	      \, \mv{I(\bi q,\omega,\Omega)} \, 
	      \rme^{-\rmi(\Omega t + \bi q\cdot\bi r)}
$ %}
with   
\eq{
	\mv{I(\bi q,\omega,\Omega)} = \mathcal{N} \int_{L^3} \rmd^3r 
         \, \Tr[\rho \, G^\dagger (\omega_-) 
	\bi E^{(-)}(\bi r) \cdot \bi E^{(+)}(\bi r) G(\omega_+)]
	\, \rme^{\rmi\bi q\cdot\bi r}. 
}
The amplitude evolves with the retarded propagator $G(\omega_+)$, and
the conjugate amplitude with the advanced propagator 
$G^\dagger (\omega_-)$. 
Here, $\omega$ is the average evolution frequency while
$\Omega$ is the frequency difference: $\omega_\pm=\omega \pm \Omega/2$. 
For small frequencies $\Omega \rightarrow 0$, the stationary regime 
(or long-time limit) 
is recovered. 

In the following, we restrict our theory to low intensity light fields, 
neglecting the non-linear response 
(\emph{i.e.} saturation) of the atomic dipole transition, by studying the
evolution of a field state containing at most one photon. 
We thus consider an initial density matrix of the form 
$\rho(0)= \rho_{\mathrm{at}} \otimes \rho_\nu$ 
where $\rho_\nu = \sum_{1,2} \rho_{1,2} \ket{1}\bra{2}$ 
describes the one-photon initial light field with 
the short-hand notation $1\equiv \bi k_1 \beps_1$. For example, 
$\rho_{1,2} = \delta_{1,i} \delta_{2,i}$ describes a pure state 
consisting of an initial plane wave $\ket{\bi k_i \beps_i}$.
Using \refeq{E}, a straightforward calculation gives the average
intensity for any incident field 
\eqlab{
\mv{I(\bi q,\omega,\Omega)} = \mathcal{N}
\sum_{1,2,3,4} \rho_{1,4} \, \mathcal{E}_2 \, \mathcal{E}_3 
(\bbeps_3\cdot\beps_2) \, \delta_{\bi k_3-\bi k_2,\bi q}\, 
\Phi(\{\bi k\beps\};\omega,\Omega) 
}{avintens1}
in terms of the \emph{intensity propagation kernel}    
\eq{
\Phi(\{\bi k\beps\};\omega,\Omega) \equiv  \mv{ 
\bra{\bi k_4\beps_4} G^\dagger (\omega_-)\ket{\bi k_3\beps_3}
\bra{\bi k_2\beps_2} G(\omega_+)\ket{\bi k_1\beps_1}} 
}
or, in operator form, $\Phi(\omega,\Omega) \equiv \mv{G^\dagger (\omega_-)
\otimes G(\omega_+)}$.  

\subsection{Bethe-Salpeter equation}

For the average amplitude, the average propagator $\mv{G(\omega)}$
had been calculated by solving the Dyson equation 
with the help of the self-energy. 
In close analogy, the intensity propagation kernel $\Phi$ 
obeys the Bethe-Salpeter equation 
\eqlab{\fl
\Phi(\omega,\Omega)=\mv{G^\dagger(\omega_-)}\otimes \mv{G(\omega_+)} 
	+ \mv{G^\dagger(\omega_-)}\otimes \mv{G(\omega_+)} U(\omega,\Omega) \,
		\Phi(\omega,\Omega). 
}{Betsal}
Here, the irreducible vertex $U(\omega,\Omega)$ contains all 
diagrams with at least one vertical
connection between the direct amplitude (upper line or ``particle
channel'') and the 
conjugate amplitude (lower line or ``hole channel'') 
 \cite{Vollhardt80}, 
\eqlab{
\intensbox{U} =  \lstep +  \crossed +   \tdiagram + \tdiagramcc + \dots 
}{Uendiagramme} 
Factors of order $N$ are omitted for brevity here. 
One defines the reducible
intensity  vertex $R(\omega,\Omega)$ by 
 \eqlab{\fl
\Phi(\omega,\Omega)=\mv{G^\dagger(\omega_-)}\otimes \mv{G(\omega_+)} 
	+ \mv{G^\dagger(\omega_-)}\otimes \mv{G(\omega_+)} R(\omega,\Omega) 
		\mv{G^\dagger(\omega_-)}\otimes \mv{G(\omega_+)}.   
}{Rdef}   
Up to a dressing by average propagators, it suffices then 
to solve the Bethe-Salpeter equation for the reducible intensity vertex, 
 \eqlab{
R(\omega,\Omega) = U(\omega,\Omega)
	+ U (\omega,\Omega) \mv{G^\dagger(\omega_-)}\otimes\mv{G(\omega_+)}
R(\omega,\Omega).    
}{RdeU} 
Diagrammatically, this equation reads  
\eqlab{
\intensbox{R} 
  = \intensbox{U} +  \intensbox{U} 
\avgintens \intensbox{R}.  
}{defUdiag}
In general, this equation cannot be solved exactly, and the
irreducible vertex $U(\omega,\Omega)$ has to be approximated. 
Any approximation of $U(\omega,\Omega)$ should be   
consistent with that of $\Sigma(\omega)$ since, physically speaking,
the depletion of in initial state is caused by scattering into
initially empty field modes. Mathematically, this consistency is
assured order by order in perturbation theory by a Ward identity
\cite{Vollhardt80}.  

\subsection{Boltzmann approximation}

In the weak-scattering regime, the irreducible vertex 
may be approximated by the first term on the right hand side of
\refeq{Uendiagramme} that represents the single-scattering 
contribution:
\eqlab{ 
U^{(1)}(\{\bi k\beps\};\omega,\Omega) 
=  N \mv{
	\bra{\bi k_4\beps_4} T^\dagger _\alpha(\omega_-) \ket{\bi k_3\beps_3}\,
	\bra{\bi k_2\beps_2}T_\alpha(\omega_+) \ket{\bi k_1\beps_1}}. 
}{UdeTdaggerT}  
This is the so-called Boltzmann approximation (or first-order smoothing 
approximation). It proves to be consistent with the independent
scattering approximation \refeq{elmatSigma} for 
the self-energy since the corresponding Ward identity 
reduces to the optical theorem \refeq{sectionefficace} for $\Omega=0$
\cite{Lagendijk96}.   
The Bethe-Salpeter equation \refeq{defUdiag} 
then acquires by iteration the familiar and simple
ladder structure  $R(\omega,\Omega)\approx
L(\omega,\Omega)$, or diagramatically: 
\eqlab{ 
\intensbox{R} \approx  
\intensbox{L} 
  = \lstep + \lstep \avgintens\lstep 
+ \lstep \avgintens \lstep \avgintens \lstep + \dots 
}{ladderseries}
Here both the direct and the conjugate amplitude are scattered by
exactly the same scatterers. Thus, it is the squared
amplitude or intensity that propagates from scatterer to scatterer,
and all interference has disappeared. This approximation leads 
for $\bi q\to 0$ to a Boltzmann 
transport equation for the intensity, justifying the Drude model for
the 
electronic conductivity or the radiative transfer equation for 
diffusive light transport \cite{Lagendijk96}.

The average single-scattering intensity \refeq{UdeTdaggerT} can be calculated
explicitly. Using expression \refeq{amplidet} for the matrix element 
of the one-atom scattering operator, one can immediately compute the 
external average and verify the total momentum conservation 
$	\langle \rme^{\rmi(\bi k_1-\bi k_2+\bi k_3-\bi k_4)\cdot \bi r_\alpha} 
  	\rangle _{\mathrm{ext}} 
 	= \delta_{\bi k_1+\bi k_3,\bi k_2+\bi k_4} $
as required by the statistical invariance under translations of the
infinite medium. 
The internal average of the squared atomic scattering operator
 \refeq{tmatrix} can be
 calculated for an arbitrary internal degeneracy using the techniques
 of irreducible tensor operators (for details, see
 \cite{Mueller01}):
\eq{\fl 
\mv{(\bbeps_4 \cdot \bi t^\dagger(\omega_-) \cdot \beps_3) 
	(\bbeps_2 \cdot \bi t (\omega_+)\cdot \beps_1)}_{\mathrm{int}}  
	= M_J \bar t(\omega_-)t(\omega_+) \; 
 	\mathcal{I}(\beps_1,\bbeps_2,\beps_3,\bbeps_4). 
} 
Here, $\mathcal{I}(\beps_1,\bbeps_2,\beps_3,\bbeps_4)$ is a vertex
 function connecting four vectors according to 
\eqlab{\fl 
 \mathcal{I}(\{\bi x\}) =  
w_1\;(\bi x_1\cdot\bi x_2) (\bi x_3\cdot\bi x_4) 
+ w_2\;(\bi x_1\cdot\bi x_3) (\bi x_2\cdot\bi x_4) 
 + w_3\;(\bi x_1\cdot\bi x_4) (\bi x_2\cdot\bi x_3) 
}{tracedet}
The weights of the three possible pairwise contractions are 
\eqlab{
w_1  = \frac{s_0-s_2}{3}, \quad  
w_2  = \frac{s_2-s_1}{2}, \quad 
w_3 = \frac{s_1+s_2}{2}
}{widesK}
where the coefficients 
\eqlab{
s_K=3(2\Je+1) \sixj{1}{1}{K}{\Jg}{\Jg}{\Je}^2
}{skde6j}
are proportional to squared $6J$-symbols or Wigner coefficients that
 are known to describe the possible re-coupling of four 
 vector operators. 
A diagrammatic representation of the atomic four-point
 vertex has been introduced,  
\eqlab{  
{\mathcal I}(\{\bi x\}) 
 \equiv 
\vertex{1\,}{\,2}{\,3}{4\,} 
  =  w_1 \ \horizontal{1\,}{\,2}{\,3}{4\,}
  +  w_2 \ \diagonal{1\,}{\,2}{\,3}{4\,}
  +  w_3 \ \vertical{1\,}{\,2}{\,3}{4\,}.
}{boxdef}
The irreducible vertex \refeq{UdeTdaggerT} is therefore given by 
 \eqlab{ 
	U^{(1)}(\{\bi k\beps\};\omega,\Omega)
 = u(\omega,\Omega) \  
 \vertex{\beps_1}{\bbeps_2}{\beps_3}{\bbeps_4} \;
 \delta_{\bi k_1+\bi k_3,\bi k_2+\bi k_4} 
}{U1}
where
$u(\omega,\Omega) =NM_J\bar t(\omega_-)t(\omega_+)$ has  the dimension
of an energy squared.

%%%%%%%%%%%%%%%%%%%%%%%%%%%%%%%%%%%%%%%%%%
\subsection{Weak localization corrections}

Finding among all possible diagrams  the dominant interference 
corrections to the Boltzmann approximation proves to be a delicate
subject \cite{Belitz94}. 
Langer and Neal \cite{Langer66} introduced the 
so-called \emph{maximally crossed} diagrams yielding an interference 
correction to the ladder terms independently of the density of scatterers:  
\eqlab{
\intensbox{C} =  \crossed + \triplecrossed +\dots 
}{crossedseries}  
These diagrams describe amplitudes that propagate
along the same scattering paths but in opposite directions. 
Their interference is responsible
for the weak localization corrections to the conductivity of electrons in
weakly disordered mesoscopic systems \cite{Bergmann84} and can be 
implemented self-consistently through $U(\omega,\Omega) \approx U^{(1)}
(\omega,\Omega)+C(\omega,\Omega)$ \cite{Vollhardt80}. This
interference also gives rise to the coherent backscattering peak
scattered from a bounded medium \cite{Akkermans88}. 
There, the interference correction is calculated 
using the heuristic prescription   
$R(\omega,\Omega)\approx L(\omega,\Omega) + C(\omega,\Omega)$. 
In sections \ref{sum.sec} and \ref{dyn.sec}, the full ladder and
crossed contributions will be calculated, preparing the ground for the
calculation of the CBS peak and weak localization corrections. 

The survival of an interference term despite an average over 
random realizations may seem 
miraculous at first glance. Closer analysis reveals that it is really
due to the fundamental symmetry of time reversal invariance. Let us
briefly recall the exact relation between ladder and crossed
diagrams. Take the first 
non-trivial term of the ladder sum \refeq{ladderseries}, the
second-order scattering contribution with matrix elements  
\eqlab{
\fl 	L_2(\{\bi k,\beps\})=
	\leftarglstep{1}{4}	 
	\makebox[0.6375em]{\avgintens}
	\rightarglstep{2}{3} 
	= \sum_{\rmu,\rmd} U^{(1)}(1,\rmu,\rmd,4) \mv{G(\rmu)}
	\mv{\overline{G}(\rmd)} U^{(1)}(\rmu,2,3,\rmd),  
}{L2diagramme}
where $\rmu\equiv\bi k_\rmu\beps_\rmu$ and $\rmd\equiv\bi
k_\rmd \beps_\rmd$ and where the dependence on $\omega,\Omega$ is
understood. 
Since each intensity vertex \refeq{U1} conserves the total momentum,
the sum also does, and the ladder
matrix element can be written 
$L_2(\{\bi k,\beps\})= \delta_{\bi k_1+\bi k_3,\bi k_2+\bi k_4} \, L_2(\bi
q,\{\beps\})$ with $\bi q=\bi k_1-\bi k_4=\bi k_2-\bi k_3$ and   
\eqlab{\fl 
L_2(\bi q,\{\beps\})= u(\omega,\Omega)^2 \sum_{\rmu,\rmd} 
	\mathcal{I}(\beps_1,\bbeps_\rmu,\beps_\rmd,\bbeps_4)
	\mv{G(\rmu)}	\mv{\overline{G}(\rmd)} 
	\mathcal{I}(\beps_\rmu,\bbeps_2,\beps_3,\bbeps_\rmd)
	\, \delta_{\bi k_{\rmd},\bi k_{\rmu}-\bi q} .
}{L2eps}
The corresponding crossed diagram 
\eqlab{
	C_2(\{\bi k,\beps\}) = 
	\argcrossed{1}{2}{3}{4} 
}{C2diag}
can be treated in the same manner, and we have to calculate 
\eqlab{\fl 
C_2(\bi q_C,\{\beps\})= u(\omega,\Omega)^2 \sum_{\rmu,\rmd} 
	\mathcal{I}(\beps_1,\bbeps_\rmu,\beps_3,\bbeps_\rmd)
	\mv{G(\rmu)}	\mv{\overline{G}(\rmd)} 
	\mathcal{I}(\beps_\rmu,\bbeps_2,\beps_\rmd,\bbeps_4)
	\, \delta_{\bi k_{\rmd},\bi q_C-\bi k_{\rmu}}
}{C2eps}
where its argument $\bi q_C=\bi k_1+\bi k_3$ is now the total
momentum. 
 
The comparison of expressions \refeq{L2eps} and \refeq{C2eps} shows
that the value of any crossed diagram (the generalization to arbitrary
scattering orders is evident) can be obtained without further
calculation from the corresponding ladder contribution by the substitutions 
\numparts
\begin{eqnarray}
\bi q  = \bi k_1-\bi k_4 \quad & 
	\mapsto  & \quad \bi q_C = \bi k_1 + \bi k_3 ,	\label{rule1.eq}\\
(\beps_3,\bbeps_4) \quad & 
	\mapsto & \quad (\bbeps_4,\beps_3),  	\label{rule2.eq}\\
	(w_2,w_3) \quad 	& 
	\mapsto & \quad  (w_3,w_2)		\label{rule3.eq}
\end{eqnarray}
\endnumparts
The first substitution rule
\refeq{rule1}, well known in the case of scalar wave scattering,  
implies that the crossed and ladder terms are equal for scattering of
plane waves ($\bi k_1=\bi k_4=\bi k$, $\bi k_2=\bi k_3=\bi k'$ such
that $\bi q=0$) in
the backwards direction $\bi k'=-\bi k$ since then also $\bi q_C=\bi
k+\bi k'=0$. The second substitution rule 
\refeq{rule2}, known for vector waves,  
restricts the equality to the channels of preserved helicity
or parallel linear polarization where $\bbeps'=\beps$. If these two
conditions in the case of isotropic point scatterers are satisfied, then
the reciprocity theorem, stemming from time-reversal
invariance, indeed justifies the equality of ladder and crossed
series.  Symbolically, the crossed diagram can be disentangled by
turning around its lower line and becomes topologically a ladder
diagram \cite{reciprocity}. 
The third
substitution rule \refeq{rule3} is new and 
arises because of the scatterers' internal
structure: by turning around the lower amplitude according to the two
previous rules, the
atomic internal vertices \refeq{boxdef} are twisted. To obtain the
correct expression, the vertical and diagonal contractions 
have to be exchanged such that $\mathcal{I}(\cdot,\cdot,\bi x_3,\bi
x_4) \mapsto \mathcal{I}(\cdot,\cdot,\bi x_4,\bi x_3)$, 
or symbolically 
\eq{
\vertex{}{}{}{} \quad\mapsto\quad 
\twistedvertex{}{}{}{} \quad\ne\quad \vertex{}{}{}{}.
}
As explained in detail in \cite{Mueller01}, the inequality
$w_2\ne w_3$ that causes a difference of crossed and ladder
contributions even for parallel polarizations 
must be attributed to the antisymmetric part $t^{(1)}$ of the
scattering operator (represented by the coefficient $s_1$ in
equation~\refeq{widesK}). 

The substitution rules \refeq{rule1}-\refeq{rule3} permit to obtain the crossed
contributions immediately from the ladder contributions or vice versa.     
In principle,  one can calculate  
diagrams of arbitrary scattering order using the above prescriptions.   
A diffusive transport for the intensity, however, 
only emerges in the limit where all ladder diagrams are summed up.

%%%%%%%%%%%%%%%%%%%%%%%%%%%%%%%%%%%%%%%%%%%%%%%%  
\section{Summation of ladder and crossed series}
\label{sum.sec}

Waves can be either scalar or vectorial, and point scatterers can be
either isotropic or anisotropic. This distinction defines four classes
of multiple scattering theories with growing complexity.     
The first case of scalar waves and isotropic 
point scatterers can be considered
well understood \cite{vanRossum99,deVries98}. 
The case of scalar waves and anisotropic point scatterers has been
solved in the framework of the radiative transfer theory \cite{Amic96}. 
The multiple scattering of electromagnetic
vector waves by point dipole scatterers, from the first approaches based on
the diffusion approximation 
\cite{Akkermans88,Stephen86,MacKintosh88} up to the exact solution
of the radiative transfer equation by the Wiener-Hopf method 
\cite{Ozrin92a,Amic97}, has kept a somewhat discouraging
appearance. Indeed, for a vector wave like light, polarization and 
direction are linked by transversality. In order to describe 
the evolution of the light intensity in three dimensions, one needs to
manipulate tensors of rank four or $9\times9$ transfer matrices. The
strategy employed in the literature consists in applying the scalar
methods to this transfer matrix, which needs to be
diagonalized in an appropriate way. The published results 
demonstrate the difficulty of the problem and the complexity of its solution.

A solution for the most difficult case of vector waves and arbitrary
scatterers is still lacking to our knowledge.      
But the scattering of light by atoms with a quantum internal structure
falls precisely into this last class of difficulties since the
internal structure couples to the polarization and the isotropic
dipole approximation $\Jg=0$ is forbidden  by definition. 
In a previous work \cite{Mueller01}, we have been able to obtain the atomic
intensity vertex $\mathcal{I}(\{\bi x\})$ for arbitrary internal
degeneracy $\Jg>0$ by a
systematic analysis in terms of irreducible operators with respect to the
rotation group. It is therefore natural to apply the same powerful
tool in order to simplify the summation of the multiple scattering series as
much as possible. In this section, we will consider the
static case $\Omega =0$, postponing the discussion of dynamic effects
to section \ref{dyn.sec}. 
    
%%%%%%%%%%%%
\subsection{Strategy of summation}

Let us first recall the solution for the case of scalar waves. This
case can be viewed as a limit of the full vector case 
by averaging over the incident
polarization $\beps$,
and summing over all final directions $\bi k'$ and polarizations
$\beps'$. The irreducible vertex \refeq{U1} then becomes the scalar
factor $u_0=\pi n \sigma/L^3\omega^2$, and the ladder series
\refeq{ladderseries} sums up to 
\eqlab{
L(q)= u_0 \left(1 + \mathcal{A}(q\ell) + \mathcal{A}(q\ell)^2 +
\dots \right) = \frac{u_0}{1-\mathcal{A}(q\ell)}.
}{scalladdersum}
Here, the momentum transfer function $\mathcal{A}$ is given by
the auto-convolution 
\eqlab{
	\mathcal{A}(q\ell) \equiv u_0    \sum_{\bi k} 
	\mv{G(k;\omega)}\mv{\overline{G}(|\bi k-\bi q|;\omega)}
}{Ascalairediag}
of the average scalar propagator \refeq{Gdiag}. 
The average real-space propagator 
$\mv{G(r;\omega)} = L^{-3}\sum_{\bi k}\mv{G(k;\omega)} \, 
\rme^{-\rmi \bi k\cdot\bi r}$ 
in scalar form reads 
\eqlab{
\mv{G(r;\omega)}= - \frac{\omega}{2\pi r} \, 
\rme^{\rmi(k-\Sigma(\omega))r} 
}{mvGderscalar}
after a 
suitable regularization of the UV-divergence and up to near-field
terms of order 
$(kr)^{-2}$ \cite{Morice}.   
Using $\Im\Sigma(\omega)=-1/2\ell$ and the Plancherel-Parseval
relation, one finds 
\eqlab{
\mathcal{A}(q\ell)=  
\frac{n \sigma}{4\pi} \int\rmd^3 r \,\frac{\rme^{-r/\ell}}{r^2} \,  
\rme^{\rmi\ps{q}{r}} 
   =  n \sigma \ell \, \frac{\arctan (q\ell)}{q\ell}.   
}{Ascalaire} 
Note that $n\sigma$ 
stems from the single scattering vertex $U^{(1)}$ whereas $\ell$ is
defined through  
the imaginary part of the self-energy. 
This distinction should be kept in mind when 
approaching the localization threshold, but in the present 
dilute regime we can use $n\sigma\ell=1$. 
$\mathcal{A}$ thus only depends on the reduced momentum
$p=q\ell$. Its small momentum behavior $
\mathcal{A}(q)=1-p^2/3+O(p^4)$ assures that the summed ladder propagator
\refeq{scalladdersum} has the usual diffusion pole at the origin, $L(q)\sim
3u_0/p^2$. 

The average intensity for vector waves 
is described in terms of four-point diagrams
that connect the incident to the scattered polarization vectors. 
The vector ladder series thus defines a ladder tensor
$\sfL_{ijkl}(\bi q)$ according to  
$L(\bi q,\{\beps\}) \equiv u(\omega,0) \,
	\eps_{1,i}\bar\eps_{2,j}\eps_{3,k}\bar\eps_{4,l}
	\,\sfL_{ijkl}(\bi q)$ 
(here and in the following, the 
summation over repeated Cartesian indices is understood). 
The factor $u(\omega,0)= 3n\sigma /
(4\pi\rho_0(\omega))$ makes the ladder tensor
dimensionless.  
The ladder series \refeq{scalladdersum} in tensor form reads 
\eqlab{
\sfL(\bi q)  = \sfI +  \sfI \,\sfA(\bi q) 
	+ \sfI\, \sfA^2(\bi q)  +\dots  
	= \sfI\, (\mathsf{1}-\sfA(\bi q) )^{-1}
}{Ltenssomme}
where the transfer tensor $\sfA$, as in \refeq{Ascalairediag}, is 
the product of the atomic single scattering intensity vertex $\sfI$ 
and the autoconvolution $\sfG$ of transverse propagators,
diagrammatically  
\eq{
\sfA_{ijkl} \equiv   (\sfG\sfI)_{ijkl} \equiv 
\leftargavgintens{i}{l}\rightarglstep{j}{k} \quad .  
}
In the multiple scattering series, the tensors are multiplied in the
``horizontal'' direction  
\eqlab{
(\sfA^2)_{ijkl} = \ \leftargavgintens{i}{l}
\rightarglstep{m}{n} \leftargavgintens{m}{n}\rightarglstep{j}{k} \quad .
}{Acarrediag}
In order to stress this important feature, we group the indices
by pairs left-right $\sfA_{il;jk}\equiv  \sfA_{ijkl}$
so that the horizontal tensor product reads explicitly 
\eqlab{
(\mathsf{AB})_{il;jk} \equiv  \sfA_{il;mn} \mathsf{B}_{mn;jk} . 
}{defproduittensoriel} 
The strategy of summation will be to diagonalize the tensors of rank
four with respect to this product.  
We will thus try to decompose the
atomic scattering vertex and the transfer tensor,   
\eqlab{
\sfI = \sum_\beta \lambda_\beta \sfT^{(\beta)} ,\qquad
\sfA =\sum_{\beta} a_\beta \sfT^{(\beta)},  
}{decomp}
in terms of suitable orthogonal projectors 
$\sfT^{(\beta)}\sfT^{(\beta')} = \delta_{\beta\beta'}
\sfT^{(\beta)}$.  
In this form, the summation of the ladder series would be 
trivial,  
\eq{
\sfL= \sum_\beta \frac{\lambda_\beta}{1-a_\beta}
\sfT^{(\beta)}.  
}
The corresponding crossed sum would then be obtained by substracting the
single-scattering term and by applying the substitution rules
\refeq{rule1}-\refeq{rule3}. This strategy proves to be successsful up
to minor complications to be discussed below. 

%%%%%%%%%%%%
\subsection{Irreducible eigenmodes of the atomic intensity vertex}

%\subsubsection{Ladder vertex:}

Let us start by analyzing the atomic ladder vertex defined by 
$\mathcal{I}(\{\beps\}) \equiv \eps_{1,i}\bar\eps_{2,j}
\eps_{3,k}\bar\eps_{4,l} \,\sfI_{ijkl}$ or, equivalently, 
\eqlab{
\sfI_{il;jk} = \vertex{i}{j}{k}{l}= w_1 \,\delta_{ij}\delta_{kl}
			+ w_2 \,\delta_{ik}\delta_{jl}
			+ w_3 \,\delta_{il}\delta_{jk} . 
}{Idewi}
The identity for the tensor product 
\refeq{defproduittensoriel}, a scalar with
respect to rotations, can be decomposed into its irreducible
components with respect to the pairs of indices $(il)$ and $(jk)$ as 
\eqlab{
\mathsf{1}_{il;jk} \equiv \horizontal{i}{j}{k}{l}= 
\delta_{ij}\delta_{kl} = \sum_K \sfT^{(K)}_{il;jk}  
}{sommeidentite}
where the scalar, antisymmetric and symmetric traceless basis tensors   
\eqlab{
\eqalign{
\sfT^{(0)}_{il;jk} & \equiv  \frac{1}{3} \delta_{il}\delta_{jk} ,\\
\sfT^{(1)}_{il;jk}& \equiv \frac{1}{2}  
	(\delta_{ij}\delta_{kl} - \delta_{ik}\delta_{jl}) , \\
\sfT^{(2)}_{il;jk} & \equiv \frac{1}{2}  (\delta_{ij}\delta_{kl} +
\delta_{ik}\delta_{jl}) -  \frac{1}{3} \delta_{il}\delta_{jk} 
}
}{TKbase}
define an algebra of orthogonal projectors,  
$\sfT^{(K)}\sfT^{(K')} = \delta_{KK'} \sfT^{(K)}$.
The irreducible cartesian components of any rank-two tensor
$M$ 
are obtained by the projection $M^{(K)}_{ij} \equiv \sfT^{(K)}_{ij;mn}
M_{mn}$. Accordingly, we define the pairwise irreducible components of
the atomic vertex as  
$\sfI^{(K,K')}\equiv \sfT^{(K)}\sfI\sfT^{(K')}$. 
Because the atomic vertex is globally invariant under rotations, only
its diagonal components $K=K'$ are non-zero, 
\eqlab{
\sfI=\sum_K \sfI^{(K,K)}= \sum_K \lambda_K \sfT^{(K)}.
}{vertexdiagonal}   
The corresponding eigenvalues are given in terms of the contraction
weights \refeq{widesK},
\eqlab{
	\lambda_0 = w_1+w_2+3w_3 = 1, \qquad
	\lambda_1 = w_1-w_2, \qquad
	\lambda_2 = w_1+w_2.      
}{lambdaKdew}
The atomic vertex function \refeq{tracedet} now reads 
\eq{
\mathcal{I}(\{\beps\})= \sum_K \lambda_K \, [\beps_1\bbeps_4]_{ij}^{(K)}
\, [\bbeps_2\beps_3]_{ij}^{(K)}. 
}
We see that the eigenvalues $\lambda_K\in[0,1]$ 
determine how faithfully the irreducible component
$[\beps_1\bbeps_4]^{(K)}$ of incident field polarizations is mapped on
average to $[\bbeps_2\beps_3]^{(K)}$. An eigenvalue $\lambda_K=1$ 
means 
perfect mapping, an eigenvalue $\lambda_K=0$ means total extinction.
Note that the scalar eigenvalue is identically $\lambda_0 = 1$ for all
$\Jg,\Je$. 
The scalar field mode being the intensity, this sum rule reflects the
conservation of energy. 
The explicit dependence of eigenvalues 
on the ground-state angular momentum $\Jg$ is given in  
\reftab{lambdaKtab} on page~\pageref{lambdaKtab.tab} and displayed in 
\refig{atvalprop} on page~\pageref{atvalprop.fig}. 
In the case of the isotropic point scatterer ($\Jg=0,\Je=1$) all eigenvalues
saturate, $\lambda_K=1$, as expected for the identity operator
\refeq{sommeidentite}. For $\Jg>0$, the non-scalar eigenvalues lie
below unity, $\lambda_{1,2}<1$. This is consistent with the physical
intuition that an initially well-polarized light beam will be
depolarized by so-called degenerate
Raman transitions between different Zeeman-sublevels 
$\ket{\Jg\mg'}\ne \ket{\Jg\mg}$.   

\begin{table}
\caption{Eigenvalues of the atomic ladder vertex, 
\refeq{lambdaKdew} and \refeq{lambdaKde6j}, as a function of the ground
state angular momentum $J$.}
\label{lambdaKtab.tab} 
\begin{indented}
\item[]\begin{tabular}{cccc}
\br
 	 & $\Je=\Jg+1$		
		&	$\Je=\Jg$		
		&	$\Je=\Jg-1$ \\
\mr
$\lambda_0$	&	$1$		& $1$ & $1$ \\ \hline
$\lambda_1$	&\rule[-1.275em]{0mm}{0mm}\rule[1.7em]{0mm}{0mm}
		$\displaystyle\frac{\Jg+2}{2(\Jg+1)} $	
		&$\displaystyle\frac{1}{2\Jg(\Jg+1)}$	
		&$\displaystyle\frac{\Jg-1}{2\Jg}$ \\\hline
$\lambda_2$	&\rule[-1.275em]{0mm}{0mm}\rule[1.7em]{0mm}{0mm}
		$\displaystyle\frac{(\Jg+2)(2\Jg+5)}{10(\Jg+1)(2\Jg+1)} $	
		&$\displaystyle\frac{4\Jg^2+4\Jg-3}{10\Jg(\Jg+1)}$	
		&$\displaystyle\frac{(\Jg-1)(2\Jg-3)}{10\Jg(2\Jg+1)}$\\
\br 
\end{tabular}
\end{indented}
\renewcommand{\arraystretch}{1.0}
\end{table}

The ``horizontal'' coefficients $\lambda_K$ can be expressed in terms
of the original coefficients $s_K$ defined in \refeq{skde6j}, 
\eqlab{
\lambda_K = \sum_{K'} (-)^{K+K'} (2K'+1)  \sixj{1}{1}{K'}{1}{1}{K}
s_{K'} . 
}{lambdaKdesK}
Indeed, the coefficients $s_K$ had been defined for the ``vertical''
coupling scheme $(ij)(kl)$, and the
irreducible recoupling of vector operators is a linear
transformation involving so-called $3nj$-symbols. The relation
\refeq{lambdaKdesK} can be recognized as a particular form of the
Biedenharn-Elliott sum rule \cite{Edmonds60} which permits to write  
\eqlab{
\lambda_K = 3(2\Je+1) \sixj{1}{1}{K}{\Je}{\Je}{\Jg}^2.   
}{lambdaKde6j}

%%%%%%%%%%%%%%%%%%%%%%%%%%%%%
%\subsubsection{Crossed vertex:} 

For the crossed series, we have to find the eigenmodes of the twisted
vertex
 \eqlab{
\sfX_{il;jk} \equiv \twistedvertex{i}{j}{k}{l} = 
	\sum_K \chi_K \sfT^{(K)}_{il;jk}.  
}{vertextordu}
By overall invariance under rotations, the same basis tensors
$\sfT^{(K)}$ as for the ladder vertex appear. The crossed
eigenvalues are obtained from the ladder eigenvalues by applying the
exchange rule \refeq{rule3}, 
\eqlab{
	\chi_0 = w_1+3w_2+w_3, \qquad
	\chi_1 = w_1-w_3,	\qquad
	\chi_2 = w_1+w_3.      
}{chiKdew}
Table~\ref{chiKtab.tab} on page~\pageref{chiKtab.tab} contains their explicit
dependence on $\Jg$, and their behaviour is displayed in
\refig{atvalprop} on page~\pageref{atvalprop.fig}.   
For the pure dipole scatterer $(\Jg=0,\Je=1)$, all ladder and crossed
eigenvalues coincide trivially $\lambda_K=\chi_K=1$. But contrary to
the ladder case, the crossed scalar eigenvalue $\chi_0$ is not fixed
by any conservation law. It indeed deviates from unity, $\chi_0<1$, as soon as
$\Jg>0$, signifying a loss of
contrast for interference corrections to the Boltzmann intensity. 
In the limit $\Je=\Jg\to\infty$, crossed and ladder eigenvalues take 
pairwise equal 
limits, signifying a perfect contrast of interference, but
non-negligeable depolarization as from a classical, anisotropic
scatterer (think of a small oriented antenna).     
A negative crossed eigenvalue, \emph{e.g.} $\chi_0=-1/3$ for 
$\Je=\Jg=1/2$, implies an even greater loss of contrast since then the
summed series behaves like   
$1/(1-\chi_K) < 1/(1-|\chi_K|)$.  

\begin{table}
\caption{Eigenvalues of the atomic crossed vertex, 
\refeq{chiKdew} and \refeq{chiKde9j}, as a function of the ground
state angular momentum $J$.}
\label{chiKtab.tab} 
\renewcommand{\arraystretch}{1.2}
\begin{indented}
\item[]
\begin{tabular}{cccc}
\br
		& $\Je=\Jg+1$		
		& $\Je=\Jg$		
		& $\Je=\Jg-1$ \\
\mr
$\chi_0$	& \rule[-1.275em]{0mm}{0mm}\rule[1.7em]{0mm}{0mm}
			$\displaystyle\frac{1}{(\Jg+1)(2\Jg+1)}$		
		& $\displaystyle\frac{\Jg^2+\Jg-1}{\Jg(\Jg+1)}$		
		& $\displaystyle\frac{1}{\Jg(2\Jg+1)}$\\\hline
$\chi_1$	& \rule[-1.275em]{0mm}{0mm}\rule[1.7em]{0mm}{0mm}
			$\displaystyle\frac{1}{2\Jg+1} $	
		& $0$	
		& $\displaystyle\frac{-1}{2\Jg+1}$ \\\hline
$\chi_2$	& \rule[-1.275em]{0mm}{0mm}\rule[1.7em]{0mm}{0mm}
			$\displaystyle\frac{6\Jg^2+12\Jg+5}{5(\Jg+1)(2\Jg+1)}$
		&$\displaystyle\frac{2\Jg^2+2\Jg+1}{5\Jg(\Jg+1)}$	
		&$\displaystyle\frac{6\Jg^2-1}{5\Jg(2\Jg+1)}$\\
\br 
\end{tabular}
\renewcommand{\arraystretch}{1.0}
\end{indented}
\end{table}

Instead of using the exchange rule $(w_2,w_3)\mapsto(w_3,w_2)$,
the crossed eigenvalues can be obtained from the ladder ones by
a partial recoupling of vectors in the vertex. 
By using the defining properties of $6j$-symbols, one finds 
\eqlab{\fl
	\chi_K = \sum_{K'}(2K'+1)\sixj{1}{1}{K'}{1}{1}{K} \lambda_{K'}
= 3(2\Je+1) \ninej{1}{\Je}{\Jg}{1}{\Jg}{\Je}{K}{1}{1} 
}{chiKde9j}
From a conceptual point of view, the expressions \refeq{lambdaKde6j}
and \refeq{chiKde9j} of the atomic vertex eigenvalues in terms of
$3nj$-symbols are fully satisfying since we have reached their most
concise, truly irreducible formulation.

\begin{figure}
\begin{center}
\includegraphics{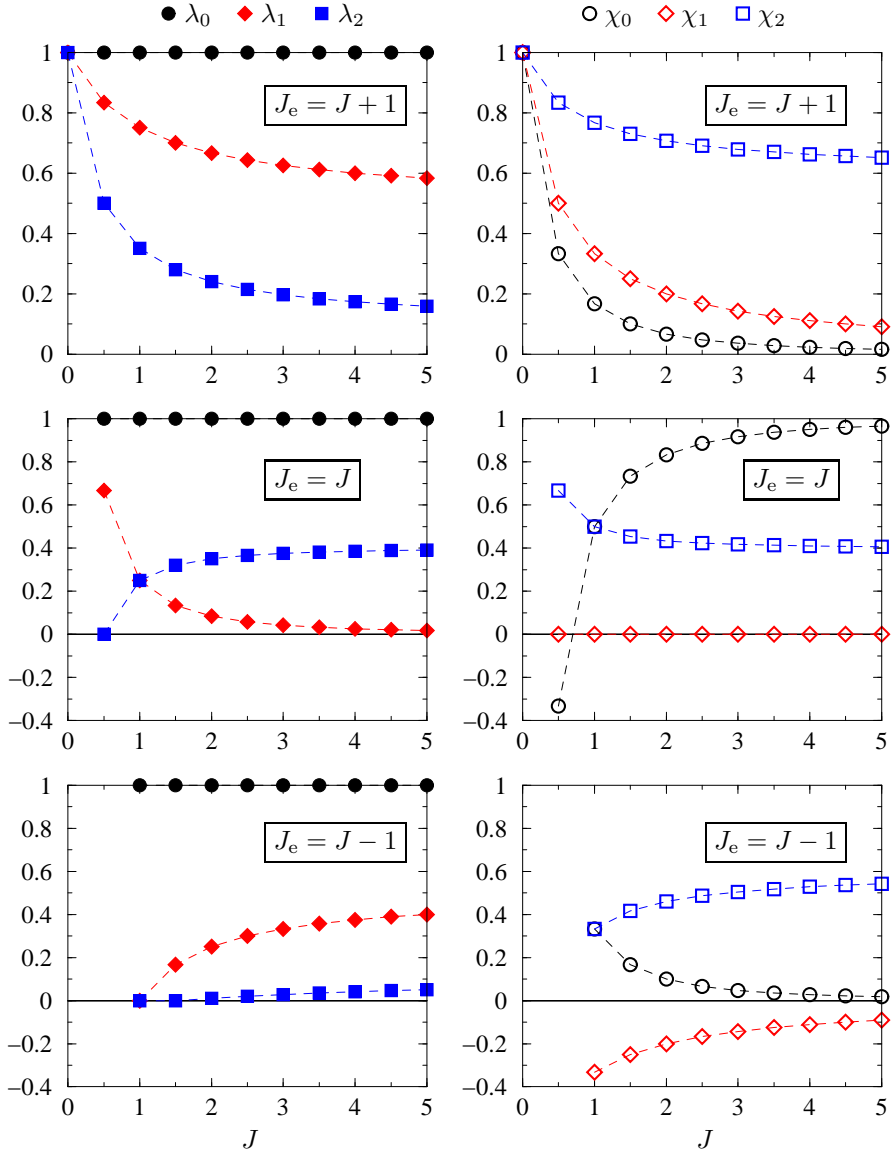}
\caption{Irreducible eigenvalues of the atomic intensity vertex as a
function of 
the ground-state angular momentum $\Jg$. 
Left: ladder eigenvalues \refeq{lambdaKde6j}. 
Right: crossed eigenvalues \refeq{chiKde9j}. 
The unit ladder eigenvalue $\lambda_0 = 1$ reflects energy
conservation. 
For $\Jg>0$, antisymmetric and 
symmetric traceless modes are not conserved $\lambda_{1,2}<1$,
reflecting depolarization. 
The crossed scalar eigenvalue $\chi_0$ plunges rapidly for 
$\Jg>0$, $\Je=\Jg+1$, implying a
drastic loss of interference contrast. 
In the semi-classical limit $\Je=J \to \infty $,  
the interference is re-established since $\chi_0 \to 1$.} 
\label{atvalprop.fig} 
\end{center}
\end{figure}

The direct product of two polarization vectors has nine independent
components.  The atomic scattering vertices therefore can be
represented by $9\times 9$ transfer matrices whose eigenvalues are  
$\lambda_K$ and $\chi_K$. Since the atomic scattering 
vertices are averages over a scalar density matrix and therefore  
invariant under rotations,  these eigenvalues are $(2K+1)$-times
degenerate (corresponding to the Clebsch-Gordan decomposition of the
direct product of representations of dimension  
 $3\times 3 = 1+3+5 $). Introducing the irreducible
components greatly simplifies a problem that at first glance defied
an analytical treatment. It is thus natural to apply this strategy to
the problem of transverse propagation as well.       

%%%%%%%%%%%%%%%%%%%%%%%%%%%%%%%%%%%%%%%%%%%%%%%%%%%%%%%%%%%%%%%%%%%%%%%
\subsection{Irreducible eigenmodes of the transverse intensity propagator}

The sum over the polarizations of an intermediate photon in scattering
diagrams like \refeq{L2diagramme} defines a transverse projector 
	$\sum_{\beps\perp\bi k}\eps_i\bar{\eps}_j=\delta_{ij}-\hat k_i
		\hat k_j$. 
The average retarded propagator for the transverse field in momentum
space therefore can be written
\eq{
\mv{G_{ij}(k;\omega)}=\frac{\delta_{ij}-\hat k_i \hat
k_j}{\omega-k-\Sigma(\omega)}.  
}
Up to near-field terms, the average real-space propagator is given by 
(\emph{cf.} \refeq{mvGderscalar})   
\eq{
\mv{G_{ij}(r;\omega)}=- \frac{\omega}{2\pi r} \, 
\rme^{\rmi(k-\Sigma(\omega))r} \Delta_{ij}   
}
where $\Delta_{ij}=\delta_{ij} - \hat r_i \hat r_j$ is the 
projector onto the plane transverse to the 
direction of propagation. 
Generalizing the scalar case \refeq{Ascalaire}, we now 
need to analyze the transverse
intensity propagator between scattering events, 
\eqlab{
\mathsf{G}_{il;jk}(\bi q) = \frac{3}{8\pi\ell}
	\int \diff^3r \, \frac{\rme^{-r/\ell}}{r^2}\,
		\Delta_{ij} \Delta_{kl}  \,\rme^{\rmi\ps{q}{r}} .
}{PKKqdePKKr}
Here, the prefactor $u(\omega,0)= 3n\sigma /
(4\pi\rho_0(\omega))$ from the atomic
vertex has been incorporated in order to manipulate a dimensionless
quantity.  
All information on the vector character
is contained in the direct product $\Delta_{ij} \Delta_{kl}$ of
transverse projectors. 
Contrary to the rotation-invariant atomic vertex,
the intensity propagator now depends on the momentum $\bi q$. 
Thus \emph{isotropic} tensor modes can no longer be sufficient,
leading to a more involved, but still exact decomposition (for
details, see \ref{appendix.sec})  
\begin{eqnarray}
\fl
\mathsf{G}_{il;jk}(\bi q)  
 = \hat s_2(p) \, P_{ij}P_{kl} + \hat s_1(p) \, 
[P_{ij}Q_{kl}+Q_{ij}P_{kl}] + 8 \hat s_3(p) \; Q_{ij}Q_{kl} \nonumber \\
  + \hat s_3(p) \; [P_{ij}P_{kl} + 2\,\mathrm{perm.}]
+ \hat s_4(p) \; [P_{ij}Q_{kl} + 5\,\mathrm{perm.}]
\label{intqDelta2.eq}
\end{eqnarray}
in terms of the projectors $Q_{ij}=\hat q_i \hat q_j$ and
$P_{ij}=\delta_{ij}-Q_{ij}$ onto the direction $\hat\bi{q}=\bi q/q$ 
and the plane perpendicular to it. 
Indeed, the projectors $Q_{ij}$ and $P_{ij}$ are the natural objects
to deal with transversality and yield the most concise expressions.  
The terms in the second line of the
right hand side of \refeq{intqDelta2} are totally symmetric with
respect to any permutation of indices which is indicated by
``$+n\,\mathrm{perm.}$''.  
The functions $\hat s_{\alpha}(p)$ are given in equations \refeq{s1s2dep}
and \refeq{s3s4s5dep} of the appendix; $p=q\ell$ is the reduced
momentum.     
The transverse projector in real space $\Delta_{ij} \Delta_{kl}$ has
the very simple structure of a 
direct product  which is obviously 
not conserved by the
Fourier transformation. In momentum space, 
the (difficult) real-space integral equation of multiple scattering
becomes a (simple) 
geometrical series of diagonal operators, but
the price to be paid is the complicated coupling between momentum and
polarization expressed by \refeq{intqDelta2}.  

Analyzing \refeq{intqDelta2} in irreducible left-right components 
$\mathsf{G}^{(K,K')}=\sfT^{(K)}\sfG\sfT^{(K')}$ as described in 
\ref{appendix.sec}, 
the transverse intensity propagator takes the form
\eqlab{
\mathsf{G}(\bi q) = \sum_{K,\alpha} g_{K\alpha} (p) \; 
	\mathsf{T}^{(K)}_\alpha(\hat\bi{q}) 
	+ \tilde g(p) \left(\mathsf{\widetilde T}^{(0,2)}(\hat\bi{q}) 
		+\mathsf{\widetilde T}^{(2,0)}(\hat\bi{q}) \right) . 
}{Gdep}
The nine $p$-dependent eigenvalues of the $9\times 9$ transfer matrix 
are partially degenerate such that only six
irreducible eigenvalues are relevant,  
\eqlab{\fl
\eqalign{
g_{00}(p)	&  \equiv \mathcal{A}(p) = \frac{\arctan(p)}{p} = 1-\frac{p^2}{3} + O(p^4),\\
g_{11}(p) &= \frac{3\left(1-\mathcal{A}(p)\right)}{2p^2} 
		= \frac{1}{2} - \frac{3p^2}{10}  + O(p^4),     \\
g_{12}(p) &= \frac{3\left(-1+(1+p^2)\mathcal{A}(p)\right)}{4p^2}
		= \frac{1}{2} - \frac{p^2}{10}  + O(p^4)  , \\
g_{20}(p) &= \frac{-9(1+p^2)+(9+12p^4+5p^4)\mathcal{A}(p)}{4p^4} 
		= \frac{7}{10} - \frac{29p^2}{210}+ O(p^4), \\
g_{21}(p) &= \frac{6+p^2-3(2+p^2-p^4)\mathcal{A}(p)}{4p^4}
		= \frac{7}{10} - \frac{13p^2}{70} + O(p^4) ,  \\
g_{22}(p) &= \frac{-3+7p^2+3(1-p^2)^2\mathcal{A}(p)}{8p^4}
		= \frac{7}{10} - \frac{23p^2}{70} + O(p^4). 
}
}{gKalphadep}

\begin{figure}
\begin{center}
\includegraphics{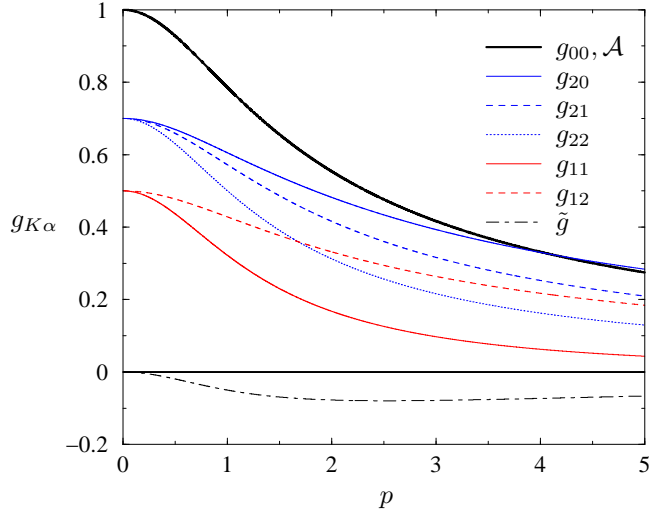}
\caption{Irreducible eigenvalues \refeq{gKalphadep} of the transverse intensity
propagator as functions of the reduced moment $p=q\ell$. 
At the isotropic limit $p=0$, the eigenvalues are $(2K+1)$-fold
degenerate, and  the coupling $\tilde g(p)$, 
equation \refeq{gtildep},
between scalar and symmetric traceless modes vanishes.} 
\label{valeurprop.fig}
\end{center}
\end{figure}

The $(2K+1)$-degenerate eigenvalues 
at the origin, $g_{K\alpha}(0)$, have been used by Akkermans
\emph{et al.}  \cite{Akkermans88} in a qualitative evaluation of
polarization effects on the coherent backscattering peak, based on the
diffusion approximation for scalar waves. The
expressions at order $p^2$ (with a small error for the coefficient of
$g_{20}$) have been found by Stephen and Cwilich
\cite[(7.4)--(4.9)]{Stephen86} and  MacKintosh and John
\cite[(4.16)]{MacKintosh88}. The above functions, noted
$\lambda_i(p)=1-g_{K\alpha}(p)$, are identical to those found by Ozrin
\cite[(3.16)]{Ozrin92a}. In a slightly modified form, they appear
also in the radiative transfer theory solved exactly by Amic, Luc and
Nieuwenhuizen \cite[(3.8),(3.15)]{Amic97}.  
But these last approaches do not make use of an explicit analysis in
terms of irreducible components, which here gives also the projectors
on the corresponding eigenmodes in terms of $P_{ij}=\delta_{ij} -\hat
q_i\hat q_j$ and $Q_{ij}=\hat q_i\hat q_j$: 
\eqlab{
\eqalign{
 \mathsf{T}^{(0)}_{0,il;jk} & \equiv \frac{1}{3}\,\delta_{il}\delta_{jk}, \\
\mathsf{T}^{(1)}_1(\hat\bi{q})_{il;jk} & 
	= \frac{1}{2}(P_{ij}P_{kl} - P_{ik}P_{jl}), \\
\mathsf{T}^{(1)}_2(\hat\bi{q})_{il;jk} & 
	= \frac{1}{2}(P_{ij}Q_{kl} + Q_{ij}P_{kl} 
			- P_{ik}Q_{jl} - Q_{ik}P_{jl}),\\
\mathsf{T}^{(2)}_0(\hat\bi{q})_{il;jk} & 
	= \frac{1}{6}(P_{il}-2Q_{il})(P_{jk}-2Q_{jk}), \\
\mathsf{T}^{(2)}_1(\hat\bi{q})_{il;jk} & 
	= \frac{1}{2}(P_{ij}Q_{kl} + Q_{ij}P_{kl} + P_{ik}Q_{jl}
	+ Q_{ik}P_{jl}), \\
\mathsf{T}^{(2)}_2(\hat\bi{q})_{il;jk} & 
	= \frac{1}{2}(P_{ij}P_{kl} + P_{ik}P_{jl}) 
		-\frac{1}{2} P_{il}P_{jk} . 
}
}{Tenseursijkl}
For the ``horizontal'' tensor product \refeq{defproduittensoriel}, the
 above tensors are orthogonal projectors, 
$\mathsf{T}^{(K)}_\alpha \mathsf{T}^{(K')}_\beta 
	= \delta_{KK'} \delta_{\alpha\beta} \mathsf{T}^{(K)}_\alpha$. 
At the limit $\bi q=0$, they recombine to give the isotropic
tensors \refeq{TKbase} 
encountered in the decomposition of the rotation-invariant atomic vertex, 
$\sum_\alpha \sfT^{(K)}_\alpha (\hat\bi{q}) = \sfT^{(K)}$. 
But this decomposition does not yield a totally diagonal
representation. 
As already pointed out by Ozrin \cite{Ozrin92a}, there exists an
irreducible coupling term between the scalar and one symmetric
traceless mode that vanishes at the origin,  
\eqlab{
 \tilde g (p)
 =\frac{\sqrt{2}[3-(3+p^2)\mathcal{A}(p)]}{4p^2} 
	= -\frac{\sqrt{2}\,p^2}{15} +O(p^4)   
}{gtildep}
with coupling tensors 
\eqlab{\fl 
\mathsf{\widetilde T}^{(0,2)}(\hat\bi{q})_{il;jk}  
	= \frac{\sqrt{2}}{6}\; \delta_{il} \; (P_{jk} - 2Q_{jk}), \qquad
\mathsf{\widetilde T}^{(2,0)}(\hat\bi{q})_{il;jk}  
	= \frac{\sqrt{2}}{6}\; (P_{il} - 2Q_{il})\;\delta_{jk}.
}{tenscouplage}
The multiplication table \refeq{tabletenscouplage} of these coupling
tensors can be found in the appendix. 
This coupling at $\bi q\ne 0$ can of course be formally diagonalized,
at the price of rather 
unwieldy expressions. At the limit of the summed multiple
scattering series, this coupling term is more simply taken care of by the
diagonalization of a $2\times2$ matrix. Furthermore, it vanishes in
the diffusion approximation.  

%%%%%%%%%%%%%%%%%%%%%%%%%%%%%%%%%%%%%%%%%%%%%%%%%
\subsection{Sum of the ladder and crossed series}

%%%%%%%%%%%%%%%
%\subsubsection{Ladder series}

The tensor $\sfA$ entering into the ladder series \refeq{Ltenssomme}
is given as the product of transverse intensity propagator and atomic
intensity vertex, 
%\eqlab{
$\mathsf{A}_L(\bi q) \equiv \mathsf{G}(\bi q)\,\mathsf{I}$. 
%}{defAL}
Knowing the respective decompositions \refeq{vertexdiagonal} and
\refeq{Gdep} into eigenmodes, the product is simply 
\eq{
 \mathsf{A}_L(\bi q) = \sum_{K,\alpha} \lambda_{K}g_{K\alpha}(p) \, 
	\mathsf{T}^{(K)}_\alpha 
	+ \lambda_2 \tilde g(p) \mathsf{\widetilde T}^{(0,2)} 
		+ \lambda_0 \tilde g(p)  \mathsf{\widetilde T}^{(2,0)}. 
} 
The message of this expression is clear: in order to generalize
the multiple scattering of light to atoms, it suffices to multiply the
vector eigenfunctions $g_{K\alpha}(p)$ by the corresponding atomic
eigenvalue $\lambda_K(\Jg,\Je)$ --- a rather trivial 
prescription once the meaning of the word 
``corresponding'' has been clarified (which constitutes the main
achievement of the present theory).   

The sum of all ladder diagrams is 
$\mathsf{L}(\bi q) = \mathsf{I}\, (\mathsf{1}-
\mathsf{A}_L(\bi q) )^{-1}$.
To invert $(\mathsf{1}-\mathsf{A}_L(\bi q))$, we try the \emph{Ansatz}
\eqlab{
(\mathsf{1}-\mathsf{A}_L)^{-1} = \sum_{K,\alpha} a_{K\alpha} \, 
	\mathsf{T}^{(K)}_\alpha 
	+ \tilde a_{2}  \mathsf{\widetilde T}^{(0,2)}
		+ \tilde a_{0} \mathsf{\widetilde T}^{(2,0)} . 
}{invansatz} 
By definition, $(\mathsf{1}-\mathsf{A}_L(\bi q)
)(\mathsf{1}-\mathsf{A}_L(\bi q)) ^{-1}=\sfid$. The identity being
$\sfid = \sum_{K,\alpha} \sfT^{(K)}_\alpha$, 
the coefficients of \refeq{invansatz}  are completely determined. For
the decoupled modes, \emph{i.e.} all except $(K=0,2,\,\alpha=0)$, the
result simply is 
$a_{K\alpha} = (1-\lambda_K g_{K\alpha})^{-1}$.
The coefficients
for the coupled modes are determined by two independent linear systems, 
\eq{
	\left(\begin{array}{c} a_{00} \\ \tilde a_{0} \end{array}\right) 
	 = \bLambda^{-1} \left(\begin{array}{c} 1 \\  0 \end{array}\right),\qquad 
	\left(\begin{array}{c} \tilde a_{2} \\ a_{20} \end{array}\right)  
	 = \bLambda^{-1} \left(\begin{array}{c}0 \\  1 \end{array}\right) .
}
Here, the $2\times 2$ coupling matrix is  
\eqlab{
 \bLambda \equiv 
	\left(\begin{array}{cc} 
	1-\lambda_0 g_{00}& -\lambda_2 \tilde g \\
	-\lambda_0\tilde g  & 1-\lambda_{2} g_{20}
	\end{array}\right), 
}{Lambda}
with determinant 
$|\Lambda|\equiv
 (1-\lambda_0 g_{00})(1-\lambda_2 g_{20}) - \lambda_0 \lambda_2\tilde g^2$. 
The coefficients of the summed series are therefore obtained by
inverting this matrix, yielding 
\eq{
a_{00} =\frac{1-\lambda_2 g_{20}}{|\Lambda|},
	\quad a_{20} =\frac{1-\lambda_0 g_{00}}{|\Lambda|},
	\quad \tilde a_{0} =\frac{\lambda_0\tilde g}{|\Lambda|},
	\quad \tilde a_{2} =\frac{\lambda_2\tilde g}{|\Lambda|}.
}
Regrouping all terms, we obtain the summed ladder series as 
\eqlab{
\mathsf{L}(\bi q) = \sum_{K,\alpha} \Lambda_{K\alpha}(p) 
	\mathsf{T}^{(K)}_\alpha(\hat\bi{q})  
	+ \widetilde \Lambda(p) \widetilde \sfT (\hat\bi{q}) 
}{Lsum} 
where the coefficients are
\eqlab{
\eqalign{
\Lambda_{00}\equiv \lambda_0 \frac{1-\lambda_{2} g_{20}}{|\Lambda|}, \qquad
\Lambda_{11}\equiv \frac{\lambda_1}{1-\lambda_{1} g_{11}},  \qquad
\Lambda_{12} \equiv \frac{\lambda_1}{1-\lambda_{1} g_{12}}, \\
\Lambda_{20}\equiv \lambda_2 \frac{1-\lambda_{0} g_{00}}{|\Lambda|}, \qquad
\Lambda_{21}\equiv \frac{\lambda_2}{1-\lambda_{2} g_{21}},  \qquad 
\Lambda_{22}\equiv \frac{\lambda_2}{1-\lambda_{2} g_{22}}, 
}
}{Lambdadep}
in terms of the atomic ladder eigenvalues $\lambda_K(J,\Je)$
defined in \refeq{lambdaKde6j} and the eigenfunctions
$g_{K\alpha}(p)$ of transverse 
propagation (equations~\refeq{gKalphadep}). 
The tensors $\sfT^{(K)}_\alpha(\hat\bi{q})$ carrying the information
about the angular dependence have been defined in \refeq{Tenseursijkl}. 
The irreducible coupling term 
\eq{
\widetilde \Lambda(p) \equiv \lambda_2 \lambda_0 \frac{\tilde g(p)}{|\Lambda|}
}
is proportional to the coupling function \refeq{gtildep} 
which vanishes as $p\to 0$. Its associated tensor is 
%\eq{
$\widetilde \sfT(\hat\bi{q}) \equiv \widetilde \sfT^{(2,0)}(\hat\bi{q}) 
+ \widetilde \sfT^{(0,2)}(\hat\bi{q})$
%} 
in terms of the left and right coupling tensors given in 
\refeq{tenscouplage}.  
 
Finally, we have to contract the external polarization vectors with
the tensors, yielding the
weights of the different field modes, 
\eqlab{
\eqalign{
t^{(K)}_{L,\alpha}(\hat\bi{q}) & \equiv
\eps_{1,i} \, \bar\eps_{2,j} \, \eps_{3,k} \, \bar\eps_{4,l}
	\,\mathsf{T}^{(K)}_{\alpha}(\hat\bi{q})_{il;jk}, \\
\tilde t_L(\hat\bi{q}) & \equiv  
\eps_{1,i} \, \bar\eps_{2,j} \, \eps_{3,k} \, \bar\eps_{4,l}
	\,\mathsf{\widetilde T}(\hat\bi{q})_{il;jk}.
}
}{contractpolarL}
Here, the populations of the different field modes are determined
by the choice of polarization, and are in particular independent
of the atomic internal structure which enters only in the
corresponding propagators $\Lambda(p;J)$. 
The summed transverse atomic ladder propagator in momentum space
reads
\eqlab{
 L(\bi q)  \equiv  u(\omega) \Big(\sum_{K,\alpha} \Lambda_{K\alpha}(p) \,
	t^{(K)}_{L,\alpha}(\hat\bi{q})
	+ \widetilde \Lambda (p)\, 
		\tilde t_L(\hat\bi{q})\Big).
}{Lsommepolar}

%%%%%%%%%%%%%%%
%\subsubsection{Crossed series}

Proceeding similarly, 
one obtains the summed crossed tensor
\eqlab{
\mathsf{C}(\bi q)= \sum_{K,\alpha} X_{K\alpha} (p)
	\mathsf{T}^{(K)}_\alpha (\hat\bi{q})  
	+ \widetilde X(p)\mathsf{\widetilde T}(\hat\bi{q}).  
}{Csum} 
The same tensors as for the ladder series appear,  
and the coefficients are obtained  
by substracting the single-scattering
contribution and replacing the atomic ladder eigenvalues
$\lambda_K$ by the corresponding crossed eigenvalues $\chi_K$, 
\eqlab{\fl
\eqalign{
X_{00}\equiv \chi_0 \left(\frac{1-\chi_{2} g_{20}}{|X|}-1\right), \qquad
X_{11}\equiv \frac{\chi_1^2 g_{11}}{1-\chi_{1} g_{11}},  \qquad
X_{12}\equiv \frac{\chi_1^2 g_{12}}{1-\chi_{1} g_{12}}, \\
X_{20}\equiv \chi_2 \left(\frac{1-\chi_{0} g_{00}}{|X|}-1\right), \qquad
X_{21}\equiv \frac{\chi_2^2 g_{21}}{1-\chi_{2} g_{21}},  \qquad 
X_{22}\equiv \frac{\chi_2^2 g_{22}}{1-\chi_{2} g_{22}}. 
}
}{Xdep}
The determinant of the crossed coupling matrix is 
$|X|\equiv
 (1-\chi_0 g_{00})(1-\chi_2 g_{20}) - \chi_0\chi_2\tilde g^2$.  
The coupling coefficient reads accordingly 
\eq{
\widetilde X(p) \equiv \chi_2 \chi_0 \frac{\tilde g(p)}{|X|}
}
The eigenvalues $\chi_K(J,\Je)$ of the twisted atomic intensity vertex
have been defined in equation~\refeq{chiKde9j}. 
The weights of the different modes are determined by contracting 
the tensors with the external polarization vectors (respecting the
substitution rule \refeq{rule2}):
\eqlab{
\eqalign{
t^{(K)}_{C,\alpha}(\hat\bi{q}) & \equiv
\eps_{1,i} \, \bar\eps_{2,j}  \, \bar\eps_{4,k}\, \eps_{3,l}
	\,\mathsf{T}^{(K)}_{\alpha}(\hat\bi{q})_{il;jk}, \\
\tilde t_C(\hat\bi{q}) & \equiv  
\eps_{1,i} \, \bar\eps_{2,j}  \, \bar\eps_{4,k}\, \eps_{3,l}
	\,\mathsf{\widetilde T}(\hat\bi{q})_{il;jk}.
}
}{contractpolarC}
As for the ladder case, the population of a field mode is determined
solely by the choice of polarization vectors (but is different from the
corresponding ladder mode unless $\bbeps_4=\beps_3$). The atomic internal
structure only enters in the corresponding propagators $X(p;J)$.  
The total summed crossed propagator therefore reads
\eqlab{
 C(\bi q)  \equiv 
	u(\omega) \Big( \sum_{K,\alpha} X_{K\alpha}(p) \,
	t^{(K)}_{C,\alpha}  (\hat\bi{q}) 
	+ \widetilde X(p)\, 
		\tilde t_C(\hat\bi{q})  \Big). 
}{Csommepolar}

%%%%%%%%%%%%%%%%%%%%%%%%%%%%%%%%%%%%%%%%%%%%%%%%%%%%%%%%%%%%%%%%%%%%%%
\section{Transport velocity, diffusion constant and relaxation times} 
\label{dyn.sec}

Let us now turn to the discussion of dynamic quantities like the
transport velocity and the diffusion constant for the different
field modes. 
In the dynamic setting $\Omega\ne 0$, the atomic ladder and crossed
vertices \refeq{vertexdiagonal} and \refeq{vertextordu} remain
unchanged. Indeed, the only 
dependence on frequency had been factorized in the prefactor
$u(\omega,\Omega)= N M_J t(\omega_+)\bar t(\omega_-)$ 
and included in the transverse ladder
kernel \refeq{PKKqdePKKr},   
\eq{
\sfG_{il;jk}(\bi q,\Omega)=u(\omega,\Omega)L^3 \int \rmd^3r
\mv{G_{ij}(\bi r;\omega_+)}
\mv{\overline{G}_{kl}(\bi r;\omega_-)} \rme^{\rmi \bi q\cdot \bi r}   
} 
where $\omega_\pm=\omega \pm \Omega/2$. Then, 
\eqlab{
\mathsf{G}_{il;jk}(\bi q,\Omega) =\mathcal{B}(\Omega)
	\frac{3}{8\pi\ell(\Omega)} 
	\int \diff^3r \, \frac{\rme^{-r/\ell(\Omega)}}{r^2}\,
		\Delta_{ij} \Delta_{kl}  \,\rme^{\rmi\ps{q}{r}}
}{GdeqOmega}
where the frequency dependence of the
propagators gives rise to an $\Omega$-dependent complex
scattering mean free path 
\eqlab{
\ell(\Omega)=\frac{1}{\rmi\Sigma(\omega_+)-\rmi\overline{\Sigma}(\omega_-)-\rmi\Omega}.  
}{elldeOmega}
The static limit is of course
$\ell(0)=\ell=-1/2\Im\Sigma(\omega)$. 
Using
$\Sigma(\omega) = N M_J t(\omega)$, the prefactor of the integral
\refeq{GdeqOmega} can be written 
\eqlab{
\mathcal{B}(\Omega) =
\frac{\Sigma(\omega_+)\overline{\Sigma}(\omega_-)}{|\Sigma(\omega)|^2}\frac{\ell(\Omega)}{\ell} 
}{B}
such that $\mathcal{B}(0) = 1 $.  Obviously,  
the tensor structure and $\bi q$-dependence in \refeq{GdeqOmega} are 
identical to the static case \refeq{PKKqdePKKr}, so that the Fourier-Laplace
transforms yield the same results, now featuring the complex mean
free path $\ell(\Omega)$.  The eigenfunctions of the transverse
propagator become 
\eqlab{
g_{K\alpha}(q,\Omega)  = \mathcal{B}(\Omega)g_{K\alpha}(p(\Omega)),
}{gKalphadyn}
where the static eigenfunctions $g_{K\alpha}(p)$, given in
\refeq{gKalphadep}, are evaluated 
at the dynamic reduced moment $p(\Omega)=q\ell(\Omega)$.   
Since the atomic eigenvalues
$\lambda_K$ and $\chi_K$ are not affected, the ladder and crossed
sums are still given by    
\refeq{Lsommepolar} and \refeq{Csommepolar}, with the dynamic
eigenfunctions \refeq{gKalphadyn} and the dynamic prefactor
$u(\omega,\Omega)$. Please note that   
these expressions are exact in $\Omega$ and $q$. 

%%%%%%%%%%%%%%%%%%%%%%%%%%%%%%%%
\subsection{Transport velocity}

In order to determine the long-time and long-distance behaviour of
transport, one habitually develops the propagator to lowest orders in
$\Omega$ and $q$. 
Since the atomic internal
structure has completely 
factorized from the frequency dependence, we exactly recover 
results that are well-known from the case of resonant scattering of scalar
waves by point particles \cite{Lagendijk96,vanRossum99}. Let us
briefly show how the main results are obtained very easily in our framework. 
The small frequency behavior of \refeq{B}, 
$\mathcal{B}(\Omega)=1+\rmi \Omega \tau_\mathrm{tr} +O(\Omega^2)$, 
defines a common transport time scale for all field modes, 
\eqlab{
\tau_\mathrm{tr} = \Im\frac{\Sigma'(\omega)}{\Sigma(\omega)} + \ell
(1-\Re \Sigma'(\omega)).
}{tautransport}
where $\Sigma'=\rmd\Sigma/\rmd\omega$. Its two contributions have simple 
physical interpretations. The first
term on the right-hand side can be traced back to the prefactor
$u(\omega,\Omega)$ 
and is simply the (Wigner) time delay  
\eq{
\tau_\mathrm{W}= \Im\frac{\Sigma'(\omega)}{\Sigma(\omega)} = 
	\frac{\rmd \phi(\omega)}{\rmd \omega}
}
where $\phi(\omega)$ is the phase of the scattering t-matrix
$t(\omega)= \Sigma(\omega)/NM_J$. The second term is the
propagation time $\ell/v_\mathrm{g}$ between consecutive scatterering
events with the
group velocity inside the effective
medium,  
$v_\mathrm{g}=\rmd \omega/\rmd
k=(1-\Re\Sigma'(\omega))^{-1}$
(remember $c=1$ and the real part of the dispersion relation, 
$k(\omega)=\omega-\Re\Sigma(\omega)$).   
The group velocity by itself looses its physical meaning in the
vicinity of a scattering resonance where extinction cannot be
neglected. 
The transport velocity 
$v_\mathrm{tr}= \ell/\tau_\mathrm{tr} =
(v_\mathrm{g}^{-1}+\tau_\mathrm{W})^{-1}$,   
however, stays causal, $v_\mathrm{tr}\le 1$, at all frequencies
$\omega$, and may decrease by orders of magnitude at resonance for
a high enough density of scatterers \cite{Albada,Lagendijk96}.  
In the present case of resonant point scatterers, the above 
expressions simplify 
remarkably since 
\eqlab{
  \frac{\Re\Sigma'(\omega)}{2\Im\Sigma(\omega)} +
\Im\frac{\Sigma'(\omega)}{\Sigma(\omega)} =\frac{1}{\Gamma}, 
}{dwelltime}
up to terms of order  $(\delta/\omega)^2,(\Gamma/\omega)^2$.  
The transport velocity then takes the simple form
$v_\mathrm{tr}= (1+ (\Gamma \ell)^{-1})^{-1}$,  
which corresponds to an inverted Lorentzian of width $\Gamma$ because
of the resonant character of the inverse scattering mean free path $\ell^{-1} 
\propto 1/(\delta^2+\Gamma^2/4)$.  
 The sum \refeq{dwelltime} has been called ``dwell time'' (in the case of non-resonant ($\Gamma\to\infty$) scattering of electrons, the
dwell time vanishes due to a Ward identity
that is invalid for scattering by frequency-dependent potentials 
\cite{Albada}).   
The total time of transport \refeq{tautransport} 
can be written as the sum of the free
propagation 
time $\ell$ between consecutive scatterers and the dwell time
$1/\Gamma$. At resonance, for example, the Wigner time delay is
$\tau_\mathrm{W} =2/\Gamma$, but the group velocity correction is 
negative since 
$\Re\Sigma'(\omega_0)/2\Im\Sigma(\omega_0)=-1/\Gamma$.   
Whereas the Wigner time delay accounts for the (possibly
large time) spent ``inside'' a resonant scattering object, the total
dwell time includes the self-consistent dressing of scatterers by the
surrounding effective medium which must not be neglected.

%%%%%%%%%%%%%%%%%%%%%%%%%%%%%%%%
\subsection{Extinction lengths}

The transversality of the light field and the atomic internal
structure are connected to the spatial variable $\bi q$. Setting
$\Omega=0$ in a first step, all propagation eigenfunctions
\refeq{gKalphadep} 
have a quadratic development in $p=q\ell$ of the form
$g_{K\alpha}(p) \approx b_K-c_{K\alpha}p^2$, whereas 
the coupling term vanishes quadratically, $\tilde g(p) = \tilde c
p^2$. At order $p^2$, the coupling between scalar and symmetric
traceless modes disappears, and the propagators \refeq{Lambdadep} 
for the transverse ladder modes behave like 
\eq{
\Lambda_{K\alpha} (p) \approx
\frac{\lambda_K}{1-\lambda_K(b_K-c_{K\alpha}p^2)} = 
\frac{c_{K\alpha}^{-1}\ell^{-2}}{q^2+\ell^{-2}_{K\alpha}}. 
}
Here appears an \emph{extinction length} 
\eq{
\ell_{K\alpha}(J) = \ell \sqrt{\frac{c_{K\alpha}}{\lambda_K^{-1}-b_K}}
}
that is responsible for an exponential
decay of the mode $K,\alpha$ in real-space (in field theory, this
corresponds to a finite particle mass $m=\ell^{-1}_{K\alpha}$). 
We notice immediately that the intensity or scalar mode $K=0$ has a
diverging extinction length since $b_0=1$ and $\lambda_0=1$ for
arbitrary internal degeneracy. This true diffusion
pole (or massless Goldstone mode) survives thanks to a fundamental
conservation law, 
the conservation of energy. All other field modes have
finite, and in fact rather short extinction lengths $\ell_{K\alpha}(J)
\lesssim \ell$, implying that the
transversality of propagation mixes well-defined polarization
modes.     
The smaller $b_K$
and $\lambda_K$, the shorter the extinction length. Thus, the atomic
internal degeneracy, responsible for $\lambda_{1,2}<1$ as soon as
$J>0$, leads to even smaller extinction lengths for
non-scalar field modes. This observation confirms the
intuitive picture that degenerate Raman transitions between different
Zeeman sublevels $\ket{Jm'} \ne \ket{Jm}$ contribute to scramble the
field polarization.  

Analogous arguments apply for the crossed propagator. Here, the atomic
ladder eigenvalues are replaced by their crossed counterparts $\chi_K$, and
the crossed extinction lengths are 
\eq{
\xi_{K\alpha}(J) = \ell \sqrt{\frac{c_{K\alpha}}{\chi_K^{-1}-b_K}}. 
}
The scalar eigenvalue $\chi_0$ is not constrained by energy
conservation and can become very small for $J>0$ 
(cf. \refig{atvalprop}). The atomic internal degeneracy
breaks the time-reversal symmetry between ladder and crossed
propagators, and leads to a rapid exponential damping of the
interference modes.       
 
Extinctions lengths of the order of or smaller than the scattering
mean free path $\ell$ imply that the evolution can no longer be
considered diffusive. Amic et al.\ \cite{Amic97} have stressed 
that the exact extinction length, defined as the inverse of the pole of the
propagator with smallest imaginary real part, can differ from the
above diffusive expression by as much as $50\%$. This difference is bound
to become even worse as the atomic eigenvalues decreases. Clearly, a
quantitative prediction for the propagation of the full vector field
must go beyond the diffusion approximation. This is especially true if
one is interested in coherent backscattering 
from a finite scattering medium,
where short paths or non-scalar field modes can become dominant. 
  
%%%%%%%%%%%%%%%%%%%%
\subsection{Diffusion constant and relaxation times}
    
Combining the developments linear in $\Omega$ and quadratic in $q$, 
the ladder propagators take the form
\eq{
\Lambda_{K\alpha}(q,\Omega) \approx
 \frac{(b_K\tau_{\mathrm{tr}})^{-1}}{-\rmi\Omega+t_K^{-1}+D_{K\alpha}q^2}. 
} 
Here appears the diffusion constant $D_{K\alpha}=
(3c_{K\alpha}/b_K)\,D_0$ for the corresponding field mode, 
simply proportional to the diffusion constant of the intensity, 
$D_0= \ell v_{\mathrm{tr}}/3$, and independent of the atomic
eigenvalues. 
Furthermore, a relaxation time has been defined,
\eq{
    t_K = \frac{\tau_{\mathrm{tr}}}{(\lambda_Kb_K)^{-1}-1}
  }
The physical meaning of this relaxation time is equivalent to that of the 
extinction lengths $ \ell_{K\alpha}$ defined above since 
$\ell_{K\alpha}^2 = D_{K\alpha} t_K$. For the scalar mode of field 
intensity, $t_0^{-1}=0$ implies a truly diffusive transport 
as required by 
local energy conservation. For the non-scalar field modes, the finite 
relaxation times $t_{1,2}$ become of the order of the transport 
time $\tau_{\mathrm{tr}}$ indicating that light transport can no longer 
be described accurately in the diffusion approximation, a tendency 
aggravated by the internal atomic degeneracy leading to $\lambda_{1,2}<1$. 
For the crossed propagator, analogous conclusions hold. The 
quantum internal structure introduces \emph{dephasing times} 
as will be discussed elsewhere \cite{MuAk}.

%%%%%%%%%%%%%%%%%%%%
\subsection{Conclusion and things to be done}

In summary, we develop a consistent theory for the multiple
scattering of photons by a dilute gas of cold atoms. The external
degrees of freedom of the atomic point scatterers are supposed to be
classical Poissonian variables. Particular attention is paid to the
internal degrees of freedom: a resonant dipole transition of
arbitrary degeneracy is treated analytically by a systematic
use of irreducible tensor operators. We sum the ladder diagrams of the
full transverse vector field, and calculate the correction of
maximally crossed diagrams.    
The internal degeneracy has no impact 
on the properties of the average amplitude (such as the scattering mean free
path) since the average over a scalar internal density matrix projects
onto the scalar component of the transition matrix. Furthermore, the
internal degeneracy is only coupled to the (static) polarization vectors and
completely factorized from the (dynamic) frequency dependence. Therefore,  
the transport velocity and the diffusion constant remain unaffected. 
However, the interference properties are strongly modified. Indeed,
the non-scalar parts of the scattering t-matrix survive in the average 
intensity
and are responsible for a decrease of contrast as soon as $J>0$. In
the diffusion approximation, we give expressions for extinction lengths or
relaxation times that depend in very simple manner on $J$. 
Atoms thus appear as an important class of
anisotropic scatterers  with intriguing interference properties
that allow a complete analytical description of light scattering.  

The present contribution deals with transport inside an infinite
scattering medium, taking full 
advantage of the statistical invariance under translations which makes
all operators diagonal in momentum representation. 
The influence of boundary conditions, relevant for coherent
backscattering or transmission experiments,
may be evaluated in the framework of the exact Wiener-Hopf method or
of the approximate method of 
images. Both approaches
deserve a detailed discussion which is beyond the scope of this paper
and will be discussed elsewhere \cite{Delande}.    
Another important issue is the influence of an external
magnetic field. Perhaps the systematic use of irreducible tensors can
help to simplify the theoretical description of light transport in 
magneto-active media \cite{vanTiggelen96}. In addition, the impact of
a splitting of the internal Zeeman degeneracy
remains to be studied, experimentally as well as theoretically.     

The transversality of the propagating light field imposes a change of
polarization in the course of multiple scattering. 
In this respect, the present theory provides a microscopic analogue of
spin-orbit coupling studied extensively in
electronic disordered systems \cite{Bergmann84}. 
But contrary to the electron case of spin $\frac{1}{2}$, 
no weak anti-localization
for the spin $1$ photon can be expected, however strong the spin-orbit
coupling.  
The scattering of photons
by atoms with internal degeneracy appears as an analogue of spin-flip
scattering of electrons by magnetic impurities \cite{Bergmann84}. 
Interestingly, we can derive exact expressions for characteristic
extinction lengths and relaxation times,  
and further experimental as well as theoretical studies can be
envisaged. The links between
optics and atomic physics on the one hand and condensed matter physics
on the other thus promise to continue to 
be an important source of inspiration to both fields.

\ack

We warmly thank A. Buchleitner, D. Delande and E. Akkermans  
for their encouragement and stimulating discussions.

%%%%%%%%%%%%%%%%%%%%%%%%%%%%%%%%%%%%%%%%%%%%%%%%%%%%%%%%%%%%%%%%%%%%
\appendix

%%%%%%%%%%%%%%%%%%%%%%%%%%%%%%%%%%%%%%%%%%%%%%%%%%%%%%%%%%%%%%%%%%%
\section{How to find the irreducible eigenmodes of the transverse
intensity propagator} 
\label{appendix.sec}

%%%%%%%%%%%%%%%
\subsection{Decomposition in real space}

Starting from the purely transverse projector in real space
 $\mathsf{G}_{il;jk}(\hat\bi{r}) \equiv \Delta_{ij} \Delta_{kl}$,
where $\Delta_{ij}=\delta_{ij}-\hat r_i\hat r_j$,     
let us first determine its ``left'' and ``right'' irreducible
components by projecting onto the isotropic basis tensors
 \refeq{TKbase}, 
\eq{
 \mathsf{G}^{(K,K')}(\hat\bi{r}) \equiv 
\sfT^{(K)} \mathsf{G}(\hat\bi{r}) \sfT^{(K')}. 
} 
The exchange symmetry $(i,j) \leftrightarrow (l,k)$ implies that the
sum of orders $K+K'$ is even since 
$ \mathsf{G}_{il;jk}^{(K,K')}(\hat\bi{r}) =  
\mathsf{G}_{li;kj}^{(K,K')}(\hat\bi{r})    
= (-1)^{K+K'} \mathsf{G}_{il;jk}^{(K,K')}(\hat\bi{r})$, 
by parity properties of the $\sfT^{(K)}$.
This condition decouples the antisymmetric from the 
symmetric modes. We find the purely scalar component
\eqlab{  
\mathsf{G}_{il;jk}^{(0,0)}(\hat\bi{r}) =
 \frac{2}{9}\,\delta_{il}\delta_{jk} ,
}{Preelscalaire}
the antisymmetric component
\eqlab{
 \mathsf{G}_{il;jk}^{(1,1)}(\hat\bi{r}) =
\frac{1}{2}(\Delta_{ij}\Delta_{kl}
	-\Delta_{ik}\Delta_{jl}), 
}{P11der}
the traceless symmetric component
\eqlab{
 \mathsf{G}_{il;jk}^{(2,2)}(\hat\bi{r}) =
	\frac{1}{2}  (\Delta_{ij}\Delta_{kl}+\Delta_{ik}\Delta_{jl})
	-\frac{1}{3} (\delta_{il}\Delta_{jk} +\delta_{jk}\Delta_{il})
 	+ \frac{2}{9} \,\delta_{il}\delta_{jk},
}{Gder2}
as well as two scalar-symmetric mixed components  
\eqlab{
  \mathsf{G}_{il;jk}^{(0,2)}(\hat\bi{r})  =
	\frac{1}{9}\,\delta_{il}\,(\Delta_{jk}-2 \hat r_j \hat r_k) ,\quad
  \mathsf{G}_{il;jk}^{(2,0)}(\hat\bi{r})  =
	\frac{1}{9}\,(\Delta_{il}-2 \hat r_i \hat r_l)\, \delta_{jk} .
}{Gdermixtes}
The irreducible components of the transverse propagator $\sfG(\bi q)$
are thus given by integrating the previous expressions according
to \refeq{PKKqdePKKr},
\eqlab{
\mathsf{G}^{(K,K')}(\bi q) = \frac{3}{2\ell}
	\int_0^\infty \diff r \, \rme^{-r/\ell} 
	\int \frac{\diff^2\hat r}{4\pi} \,\mathsf{G}^{(K,K')}(\hat\bi{r}) \,
	\rme^{\rmi\ps{q}{r}}.  
}{PKKint}
We introduce the  
short-hand notation $\mathsf{G}^{(K,K')}(\bi q) \equiv \int_{\bi q}
\mathsf{G}^{(K,K')}(\hat\bi{r})$ for this Fourier-Laplace transform. 
The angular integral of the expressions
\refeq{Preelscalaire}-\refeq{Gdermixtes} depends on the number of
times that the components $\hat r_i$ of the unit vector appear: we
distinguish scalar terms of type $\delta_{ij}\delta_{kl}$, quadratic
terms of type $\hat r_i \hat r_j \delta_{kl}$ and a quaternary term $\hat
r_i \hat r_j \hat r_k \hat r_l$. 

%%%%%%%%%%%%%%%%%%%%%%%%%%%%%%%
\subsection{Scalar component}

The angular integral of the scalar
terms is trivial, of course, and the result of the Fourier-Laplace
transform is 
\eqlab{
\int_{\bi q} 1 \equiv  \hat s_0(p)
	=\frac{3}{2}\mathcal{A}(q\ell),
}{intq1}
proportional to the scalar transfer function $\mathcal{A}(p) =
\arctan(p)/p$. The scalar component of the transverse propagator
therefore is simply given by
\eqlab{
\mathsf{G}^{(0,0)}(q) =\mathcal{A}(q\ell)\, \sfT^{(0)} 
}{G00}
 using the definition \refeq{TKbase} of the scalar basis tensor
$\sfT^{(0)}$.  
Naturally, the scalar component only depends on the transfer function
already known from the purely scalar case.

%%%%%%%%%%%%%%%%
\subsection{Antisymmetric components}

The angular integral of the quadratic terms is easily solved 
using a generating function argument,
\eqlab{
\int \frac{\diff^2\hat r}{4\pi} \,\hat r_i\hat r_j
	\rme^{\rmi\ps{q}{r}} =
-\frac{1}{r^2}\frac{\partial^2}{\partial q_i\partial q_j}
\frac{\sin(qr)}{qr} . 
}{intangq2}
Call for brevity
$s(x) \equiv \sin(x)/x$,  $x=|\bi x|$, $\bi x \equiv r \bi q$,
$\partial_i\equiv \partial/\partial x_i$, and $s'(x)=(\rmd/\rmd
x)s(x)$ so that \refeq{intangq2}
evaluates to 
\eq{
-\partial^2_{ij} s(x) = - s''(x)\, \hat q_i\hat q_j  - \frac{s'(x)}{x} \,
	(\delta_{ij} - \hat q_i \hat q_j).
}
The derivation (or equivalent angular integration)
generates the
quadratic terms $\hat q_i \hat q_j$ that could be expected, but also a
new term proportional to $\delta_{ij}$. Define the
two orthogonal projectors on the subspaces
parallel and perpendicular to $\hat\bi{q}$,  
$Q_{ij} \equiv \hat q_i \hat q_j$ and $P_{ij} \equiv \delta_{ij} -\hat
q_i\hat q_j$. They add up to the identity $\delta_{ij}$ and satisfy
\eqlab{
	P_{im} P_{mj} = P_{ij},\quad Q_{im} Q_{mj} = Q_{ij} ,
 	\quad P_{im} Q_{mj} = Q_{im} P_{mj} = 0 . 
}{PQprojorth} 
The Fourier-Laplace integral of quadratic terms is thus given by
\eqlab{	
\int_{\bi q} \hat r_i\hat r_j  = \hat s_1(p) \;P_{ij} +\hat s_2(p) \;Q_{ij}   
}{intq2}
where the coefficients are 
\eqlab{
\eqalign{
\hat s_1(p) & \equiv -  \frac{3}{2p} \int_0^\infty\diff x \, \rme^{-x/p}\,
			\frac{s'(x)}{x} 
= - \frac{3[1-(1+p^2)\mathcal{A}(p)]}{4p^2},   \\
\hat s_2(p) & \equiv  -  \frac{3}{2p} \int_0^\infty\diff x \, \rme^{-x/p} \,s''(x) 
	= \frac{3[1-\mathcal{A}(p)]}{2p^2}.
}
}{s1s2dep}
The three functions $\hat s_0(p)$ (defined in \refeq{intq1}), 
$\hat s_1(p)$ and $\hat s_2(p)$ are 
not independent. Indeed, the contraction of indices in the quadratic term 
\refeq{intq2} must reproduce the scalar integral \refeq{intq1}, so that 
$\hat s_0(p) = 2 \hat s_1(p) + \hat s_2(p)$. 

The purely antisymmetric component of the transverse propagator
\refeq{P11der} contains only scalar and quadratic terms, so that the
previous results permit to write 
\eqlab{
\mathsf{G}^{(1,1)}(\bi q) = g_{11}(p) \, \sfT^{(1)}_1
(\hat\bi{q}) + g_{12}(p) \, \sfT^{(1)}_2 (\hat\bi{q}). 
}{G1deq}
The projectors depend on the direction $\hat\bi{q}$,
\eq{
\eqalign{
\sfT^{(1)}_1(\hat\bi{q})_{il;jk} & 
	\equiv \frac{1}{2}(P_{ij}P_{kl} - P_{ik}P_{jl}), \\
\sfT^{(1)}_2(\hat\bi{q})_{il;jk} & 
	\equiv \frac{1}{2}(P_{ij}Q_{kl} + Q_{ij}P_{kl} - P_{ik}Q_{jl} - Q_{ik}P_{jl}).
}
}
Thanks to the relations \refeq{PQprojorth}, they are
indeed orthogonal projectors for the horizontal tensor product
\refeq{defproduittensoriel}, 
$\sfT^{(1)}_\alpha \sfT^{(1)}_\beta 
  = 	\delta_{\alpha\beta} \,\sfT^{(1)}_\alpha$. 
By construction, they are purely antisymmetric from left and right, so
that the antisymmetriser $\sfT^{(1)}$ commutes with them, 
$\sfT^{(1)} \sfT^{(1)}_\alpha 
	= \sfT^{(1)}_\alpha 
	= \sfT^{(1)}_\alpha \sfT^{(1)}$.
The eigenvalues of the decomposition \refeq{G1deq} depend on the
reduced momentum $p=q\ell$,  
\eqlab{
\eqalign{
g_{11}(p) & \equiv \hat s_2(p) 
		= \frac{3[1-\mathcal{A}(p)]}{2p^2} ,    \\
g_{12}(p) & \equiv \frac{\hat s_0(p)-\hat s_2(p)}{2} 
		= \frac{3[-1+(1+p^2)\mathcal{A}(p)]}{4p^2} . 
}
}{a1dep}
It is instructive to consider the limit of zero momentum
$\bi q\to 0$, where all dependence on the direction $\hat\bi{q}$ must
vanish. Both eigenfunctions 
\eq{
g_{11}(p)  = \frac{1}{2} - \frac{3p^2}{10}  + O(p^4),     \qquad
g_{12}(p)  = \frac{1}{2} - \frac{p^2}{10}  + O(p^4),  
}
have the common limit $g_{1\alpha}(0)=\frac{1}{2}$, so that the two
$\hat{\bi q}$-dependent tensors recombine to the isotropic antisymmetric
projector,  
\eq{
\sfT^{(1)}_1 (\hat\bi{q})_{il;jk} + \sfT^{(1)}_2
(\hat\bi{q})_{il;jk} =  \frac{1}{2}(\delta_{ij}\delta_{kl}
-\delta_{ik}\delta_{jl}) = \sfT^{(1)}_{il;jk}.  
}
The antisymmetric mode $K=1$ must have three eigenvalues that 
are degenerate at zero momentum where indeed $g_{1\alpha}(0)=\frac{1}{2}$. At
finite momentum $p>0$, the degeneracy is partially lifted, and the two
distinct eigenvalues $g_{11}(p)$ and 
$g_{12}(p)$ appear. 
Ozrin obtains the same eigenfunctions 
(noted $1-\lambda_9(p)$ and $1-\lambda_5(p)$, respectively, \cite[(3.16)]{Ozrin92b}) 
and shows that the remaining twofold degeneracy is carried by the
function $g_{12}(p)$. Note that we  express the propagator in
Cartesian components $\sfG^{(K,K')}_{il;jk}$ 
and not in its re-coupled spherical components 
$\sfG^{(K,K')}_{m;m'}=\sum_{pqrs}\cg{Km}{11pq}\cg{K'm'}{11rs}
	\sfG^{(K,K')}_{pq;rs}$
which turn
out not to diagonalize the propagator and therefore are not well
adapted for our purposes.

%%%%%%%%%%%%%%%%
\subsection{Symmetric traceless components}
  
For the symmetric traceless components \refeq{Gder2} of the
transverse propagator the  angular integral over a quaternary term, 
\eq{
\int \frac{\diff^2\hat r}{4\pi} \,\hat r_i\hat r_j \hat r_k\hat
r_l \, \rme^{\rmi\ps{q}{r}} =
\partial^4_{ijkl} s(x)
} 
has to be calculated, giving 
\begin{equation}
\fl 
\partial^4_{ijkl} s(x)  = 
	\frac{1}{x} \left(\frac{s'}{x}\right)'
		(P_{ij} P_{kl} + 2 \,\mathrm{perm.})
	+ \left(\frac{s'}{x}\right)''
		(P_{ij} Q_{kl} + 5 \,\mathrm{perm.})
	+ s''''(x) \, Q_{ij}Q_{kl}.
\end{equation}
The result must be totally symmetric with respect to any permutation of
indices which is indicated by ``$+n\,\mathrm{perm.}$''. The
total Fourier-Laplace transform then takes the following
form, 
\eqlab{\fl 
\int_{\bi q}  \hat r_i\hat r_j \hat r_k\hat r_l  = 
	\hat s_3(p) \; (P_{ij}P_{kl} + 2\,\mathrm{perm.}) 
	+ \hat s_4(p) \; (P_{ij}Q_{kl} + 5\,\mathrm{perm.}) + \hat s_5(p) \; Q_{ij}Q_{kl}, 
}{intq4}
with the $p$-dependent coefficients 
\eqlab{\fl 
\eqalign{
\hat s_3(p) & \equiv   \frac{3}{2p} \int_0^\infty\diff x \, \rme^{-x/p}\,
		\frac{1}{x} \left(\frac{s'}{x}\right)' 
=  \frac{-(3+5p^2) + 3 (1+p^2)^2\mathcal{A}(p)}{16p^4},   \\
\hat s_4(p) & \equiv    \frac{3}{2p} \int_0^\infty \diff x \, \rme^{-x/p} \,
		 \left(\frac{s'}{x}\right)''
	= \frac{3+2p^2-3(1+p^2)\mathcal{A}(p)}{4p^4} , \\
\hat s_5(p) & \equiv    \frac{3}{2p} \int_0^\infty \diff x \, \rme^{-x/p} \, s''''(x)
	= \frac{-3 + p^2 + 3\mathcal{A}(p)}{2p^4} .
}
}{s3s4s5dep} 
Again, these functions are not independent because a contraction of
indices in \refeq{intq4} must reduce to \refeq{intq2}, implying 
$\hat s_1(p) =  4\hat s_3(p) + \hat s_4(p)$ and 
$\hat s_2(p) =  2\hat s_4(p) + \hat s_5(p)$. 

Knowing \refeq{G1deq}, the symmetric traceless
components of the transverse propagator can be predicted to have the form  
\eqlab{
\mathsf{G}^{(2,2)}(\bi q) = g_{20}(p) \, \mathsf{T}^{(2)}_0
(\hat\bi{q}) + g_{21}(p) \, \mathsf{T}^{(2)}_1 (\hat\bi{q}) 
+ g_{22}(p) \, \mathsf{T}^{(2)}_2 (\hat\bi{q}).  
}{G2deq}
Indeed, for zero momentum $\bi q=0$, the three anisotropic tensors
$\sfT^{(2)}_{\alpha} (\hat\bi{q})$ must recombine to the isotropic
projector 
$\mathsf{T}^{(2)}_{il;jk} = \frac{1}{2}(\delta_{ij}\delta_{kl}
	+ \delta_{ik}\delta_{jl}) -\frac{1}{3} \delta_{il}\delta_{jk}$.
At non-zero momentum, the degeneracy is lifted, and each identity is
replaced by $\delta_{rs} = P_{rs} + Q_{rs}$, such that 
\eq{\fl 
\eqalign{
 \mathsf{T}^{(2)}_{il;jk} & = \frac{2}{3}\, Q_{il}Q_{jk} \\
	& \quad + \frac{1}{2}(P_{ij}Q_{kl}+Q_{ij}P_{kl} 
	+ P_{ik}Q_{jl}+ Q_{ik}P_{jl}) 
	-\frac{1}{3} (P_{il}Q_{jk}+Q_{il}P_{jk})\\
	&\quad +\frac{1}{2}(P_{ij}P_{kl}
	+ P_{ik}P_{jl}) -\frac{1}{3} P_{il}P_{jk}
          }
}
By applying the projector $\mathsf{T}^{(2)}$ to the terms of order $0,1,2$ in $P_{ij}$ separately, we find the
following traceless symmetric expressions, 
\eqlab{
\eqalign{
\mathsf{T}^{(2)}_0(\hat\bi{q})_{il;jk} & 
	\equiv \frac{1}{6}(2Q_{il}-P_{il})(2Q_{jk}-P_{jk}), \\
\mathsf{T}^{(2)}_1(\hat\bi{q})_{il;jk} & 
	\equiv \frac{1}{2}(P_{ij}Q_{kl} + Q_{ij}P_{kl} + P_{ik}Q_{jl}
	+ Q_{ik}P_{jl}), \\
\mathsf{T}^{(2)}_2(\hat\bi{q})_{il;jk} & 
	\equiv \frac{1}{2}(P_{ij}P_{kl} + P_{ik}P_{jl}) 
		-\frac{1}{2} P_{il}P_{jk}.  
}
}{tensprop22}
Again, these tensors are orthogonal projectors, 
$\mathsf{T}^{(2)}_\alpha \mathsf{T}^{(2)}_\beta  = 
	\delta_{\alpha\beta} \, \mathsf{T}^{(2)}_\alpha$. 
By construction, their partial left and right traces vanish, 
$(\mathsf{T}^{(2)}_\alpha)_{mm;jk} 
	= (\mathsf{T}^{(2)}_\alpha)_{il;nn}
	=0$.
The corresponding eigenfunctions are 
\eqlab{\fl 
\eqalign{
g_{20}(p) & \equiv -\frac{2}{3} \hat s_0(p) + \hat s_2(p) + 12 \hat s_3 (p) 
	= \frac{-9(1+p^2)+(9+12p^4+5p^4)\mathcal{A}(p)}{4p^4},    \\
g_{21}(p) & \equiv \frac{3}{2}\hat s_0(p)- \frac{3}{2}\hat s_2(p)-8\hat s_3(p) 
	= \frac{6+p^2-3(2+p^2-p^4)\mathcal{A}(p)}{4p^4},  \\
g_{22}(p) & \equiv \hat s_2(p) + 2 \hat s_3(p) 
		= \frac{-3+7p^2+3(1-p^2)^2\mathcal{A}(p)}{8p^4}, 
}
}{a2dep}
confirming the expressions 
for $\lambda_{2}=1-g_{20}$, $\lambda_{4,6}=1-g_{21}$ and 
$\lambda_{3,8}=1-g_{22}$ of Ozrin \cite[eq. (3.16)]{Ozrin92b}.  
We recall their behavior close to the origin, 
\eqlab{
\eqalign{
g_{20}(p) & = \frac{7}{10} - \frac{29p^2}{210}+ O(p^4),    \\
g_{21}(p) & = \frac{7}{10} - \frac{13p^2}{70} + O(p^4),    \\
g_{22}(p) & = \frac{7}{10} - \frac{23p^2}{70} + O(p^4).
}
}{a2develop}  

%%%%%%%%%%%%%%%%%
\subsection{Mixed symmetric modes}

Finally, the mixed scalar-symmetric modes of the transverse propagator are
\eq{
\mathsf{G}^{(0,2)}(\bi q)= \tilde g (p)
	\mathsf{\widetilde T}^{(0,2)}(\hat\bi{q}), \qquad
\mathsf{G}^{(2,0)}(\bi q) = \tilde g (p)
	\mathsf{\widetilde T}^{(2,0)}(\hat\bi{q}),
}
in terms of the mixed tensors 
\eqlab{
\fl \mathsf{\widetilde T}^{(0,2)}(\hat\bi{q})_{il;jk} 
	\equiv \frac{\sqrt{2}}{6}\; \delta_{il} \; (P_{jk} - 2Q_{jk}), \qquad
\mathsf{\widetilde T}^{(2,0)}(\hat\bi{q})_{il;jk}  
	\equiv \frac{\sqrt{2}}{6}\; (P_{il} - 2Q_{il}) \; \delta_{jk} .
}{tenseurs20}
These tensors are no longer projectors; 
their multiplication table is 
\eqlab{
\eqalign{
 \mathsf{\widetilde T}^{(K,K')}\mathsf{\widetilde T}^{(K'',K''')} 
	& = \delta_{K'K''}   \, \mathsf{T}^{(K)}_0, \\
 \mathsf{\widetilde T}^{(K,K')} \mathsf{T}^{(K'')}_\alpha 
	& = \delta_{K'K''} \, \delta_{\alpha0} \, \mathsf{\widetilde
	T}^{(K,K')}, \\
\mathsf{T}^{(K)}_\alpha  \mathsf{\widetilde T}^{(K',K'')} 
	& = \delta_{KK'} \, \delta_{\alpha0} \, \mathsf{\widetilde
	T}^{(K',K'')} . 
}
}{tabletenscouplage}
The coupling function is given by 
\eqlab{
 \tilde g (p) \equiv - \frac{\sqrt{2}[ \hat s_0(p) - 3 \hat s_2(p)]}{6} 
		= - \frac{\sqrt{2}[-3+(3+p^2)\mathcal{A}(p)]}{4p^2}, 
}{bdep}
and vanishes quadratically at the origin, 
\eq{
\tilde g(p) = -\frac{\sqrt{2}\,p^2}{15} +O(p^4).    
}

\end{document}